Universität des Saarlandes

Fachrichtung 4.7 Allgemeine Linguistik

Computerlinguistik

# Assessing Wikipedia-Based Cross-Language Retrieval Models

Submitted for the degree of M.Sc.

Supervisor: Dietrich Klakow

Benjamin Roth

beroth@CoLi.Uni-SB.de

Saarbrücken, 14 November 2009

**Eidesstattliche Erklärung**

Hiermit erkläre ich, dass ich die vorliegende Arbeit selbständig verfasst und keine anderen als die angegebenen Hilfsmittel und Quellen verwendet habe.

Saarbrücken, den 14.11.2009

Benjamin Roth

# Contents

# 1 Introduction

The task of cross-language information retrieval is to find documents relevant to a query, where query and documents do not belong to the same language. While many systems just translate either queries or documents and then perform monolingual retrieval, better performing or less ressource intensive approaches might integrate translation knowledge as an integral part of the retrieval model. Our focus will lie on integrating knowledge from the multilingual, freely available on-line encyclopedia Wikipedia[1] as a cross-language bridge for retrieval.

Dimensionality reduction techniques have traditionally been of interest for information retrieval as a means of mitigating the word mismatch problem. Cross-language information retrieval can be viewed as an extreme case of word mismatch. More general than dimensionality reduction, the term *concept model* is used to denote a mapping from the word space to another representation. Such a representation may, for example, be obtained by matrix approximation [Deerwester et al., 1990], by probabilistic inference [Steyvers and Griffiths, 2007] or by techniques making use of the conceptual structure of corpora such as Wikipedia [Gabrilovich and Markovitch, 2007].

While some work has been done on multilingual concept modeling [Dumais et al., 1997, Ni et al., 2009, Mimno et al., 2009, Potthast et al., 2008, Sorg and Cimiano, 2008, 2009], most often the focus is on one method and a comparison with other methods is missing or a particularly simple instantiation is used as a base-line. One reason for this might be that concept models require adaptations for multilinguality that do not seem to be easily implemented. We will show that the adaptations can in fact be minimal and on the data side only. Another question that has not been investigated so far is how different multilingual concept models interact with each other and how they can be combined with bag-of-word models, an approach that is standard for the monolingual case.

The subsequent parts of this thesis are structured as follows: In chapter 2 we give a broad survey over retrieval methods and knowledge sources used for cross-language retrieval. In chapter 3 we discuss three concept models included in our experiments and especially focus on their multilingual adaptations. In chapter 4 we report experiments on three corpora. We show that for one of our adaptations much better results can be observed than reported so far. In chapter 5 we explore whether machine translation output can altogether be replaced by information from Wikipedia and show encouraging results. In chapter 6 we summarize our findings and point to promising directions for further research.

---

[1] http://wikipedia.org/

# 2 Related Work and Ressources

## 2.1 Retrieval Methods

### 2.1.1 Vector Space Retrieval

Classical methods of IR follow the vector space model: documents and queries are vectors while similarity is measured by a distance funtion, most often by taking the cosine between their angles. The most simple models of that kind work directly on bag-of-word vectors and are prone to sparseness and missing word overlap, some backing-off can be achieved by dimensionality reduction techniques such as Latent Semantic Analysis [Deerwester et al., 1990] or term expansion as in the generalized vector space model [Wong et al., 1985]. In practice, many different features associated with documents can be expressed as vectors to form the basis of retrieval.

While vector space retrieval models are set in a simple and precise mathematical theory, which make them immediately accessible to many other tasks such as document clustering or classification, the connection with the actual retrieval process is not very clearly represented, especially after applying term weighting functions like the empirically well-working tf.idf scheme [Salton and Buckley, 1987]. Similar starting points and theoretical properties exhibit tuned ranking functions like the popular Okapi-BM25 family [Robertson et al., 1995], which are still achieving state-of-the-art results [Armstrong et al., 2009].

### 2.1.2 Language Modeling Retrieval

Language model based retrieval is a newer paradigm [Ponte and Croft, 1998], where usually documents are treated as providing probabilistic models of generating the queries, and the document that provides the best model is regarded as most relevant. Language model retrieval systems are among the most popular today, as they model the retrieval process in a statistically sound and consistent way and, even in their most simple forms, exhibit performance comparable to or even better than fine tuned and tweaked vector space models. See [Liu and Croft, 2004] for an excellent overview.

Usually the assumption is made that the probabilities of the query words are independent of each other conditioned on the model provided by the current document. The probabilistic models obtained from the documents $D$ are in the simplest case the maximum likelihood estimates of the unigram word probabilities $P(w|D)$. These estimates are zero for unseen words, which makes the probability of the query

$$P(Q) = \prod_{w \in Q} P(w|D)^{n(w,Q)}$$

zero if one query word has no evidence in the document.

Two different strategies are usually applied to overcome this problem. The first is to interpolate the document estimates with a naturally smoother distribution estimated from a big document collection [Hiemstra, 1998, Miller et al., 1999, Song and Croft, 1999]. The second is to smooth the estimates of either distribution by one of the many techniques originally developed for speech recognition [Chen and Goodman, 1999, Zhai and Lafferty, 2004].

### 2.1.3 Latent Variable Models

Latent variable modeling lies somewhere between vector space and language modeling retrieval: the underlying principle is to fit a probabilistic model to word–document counts, word and documents are assumed to be conditionally independent on a set (of chosen size) of latent topics. Since such models perform a sort of dimensionality reduction (for every document or query they store only the distribution of the topics) parallels can be drawn to LSA. On the other hand, in their very nature they are probabilistic, even parts can be identified that correspond directly to generative language models. Indeed, both using the topic distributions (or other parameter vectors) directly for distance comparison (as in [Hofmann, 1999, 2001, Blei et al., 2003]) as well as using them as components in a generative model (as in [Azzopardi et al., 2004, Wei and Croft, 2006]) are common practices. The two most prominent latent variable models are probabilistic Latent Semantic Analysis [Hofmann, 2001], which implements the ideas of variable modeling in a straightforward way but does not define inference on documents not present in the training collection, and, developed some years later, Latent Dirichlet Allocation [Blei et al., 2003], which defines a fully generative model and is also favorable in terms of scalability [Wang et al., 2009].

Latent variable models applied for cross-language retrieval play a central role in this thesis. Probabilistic Latent Semantic Analysis and Latent Dirichlet Allocation are described in more detail in chapters 3.1 and 3.2 respectively.

## 2.2 Using Cross-language Knowledge Sources

Knowledge sources to provide a connection between two or more languages can be manually edited lexica, parallel corpora, probabilistic lexica extracted from them and comparable corpora. Parallel corpora, such as the Europarl corpus of EU-Parliament proceedings [Koehn, 2005], provide translations of the same text in different languages, corresponding phrase pairs can be extracted automatically [Koehn et al., 2003]. Slightly different are comparable corpora, such as Wikipedia. They contain texts that have counterparts in other languages with which they are linked by dealing with the same topic. Here, it cannot in general be assumed to find corresponding phrase pairs, and it is an interesting research question how much this kind of information can contribute to effective retrieval. Also, comparable corpora are much easier to acquire, cheaper and available in more languages than parallel corpora.

### 2.2.1 Comparable Corpora

Comparable corpora provide a promising knowledge source for information retrieval. As comparable corpora just provide connections on text level, the information provided by them should be appropriately captured by dimensionality reduction techniques aiming at modeling and smoothing document-term statistics. Indeed, dimensionality reduction for cross-language retrieval has been dealt with in several theoretical frameworks and is still a field of ongoing research. We will now present work relevant to using comparable corpora and outline strengths and weaknesses of them as well as possible extensions and modifications.

The first approach [Dumais et al., 1997] to using comparable corpora and dimensionality reduction was done with Latent Semantic Analysis (LSA), a technique making use of lower-rank matrix approximation by singular value decomposition. As an effect, terms and documents are represented by lower-dimensional vectors, making them topically comparable [Deerwester et al., 1990]. For cross-lingual LSA, parallel or comparable documents are embedded and the weighted sum of embedded term vectors represents queries and documents on either side. While the method works well and is still used as a baseline [Cimiano et al., 2009], it lacks, as does monolingual LSA, a probabilistic interpretation and is hardly scalable, which becomes a major problem when it comes to embedding gigabyte-sized corpora such as Wikipedia.

A more modern approach could be based on probabilistic Latent Semantic Analysis (pLSA Hofmann [2001]). The underlying assumption in this kind of model is that words and documents are independent conditioned on an unobserved class variable $Z$ that can take on $K$ different values $z_k \in \{z_1 \cdots z_K\}$. Therefore, in this model the following factorization of the joint probablity of a word $w_i$ occurring in document $d_j$ is possible:

$$P(w_i, d_j) = \sum_{z_k} P(w_i|z_k)P(w_j|z_k)P(z_k)$$

Another, equivalent factorization is

$$P(w_i, d_j) = P(d_j) \sum_{z_k} P(w_i|z_k)P(z_k|d_j)$$

An instance of the EM-Algorithm is used to estimate an optimal factorization. The result can be used in a probabilistic fashion or by taking the cosine of the distribution vectors $p(Z|d_j)$. No work is known to us that applies pLSA to embed multilingual documents for IR. In chapter 3.1.4 we present an adaptation of pLSA for cross-lingual IR that includes a language variable $L$ taking on a language value $l_s$ and quantifying the amount of text in a particular language of a parallel document. Drawbacks of this method are theoretic concerns that the model is not fully generative and does therefore not include the possibility of estimating unseen documents (although ad-hoc techniques such as folding-in are used for that in practice). Another concern is that to us no large-scale implementations are known that showed positive scalability properties of pLSA to gigabyte sized collections like Wikipedia.

Latent Dirichlet Allocation (LDA) overcomes some of the theoretical and practical limitations of LSA and pLSA. Still the likelihood of document–word statistics is maximized, but the documents themselves (not only the terms within them) are assumed to be generated by a probabilistic process. This model allows to overcome overfitting to some degree and defines inference on unseen data in a theoretically sound way. Secondly, approximate inference by sampling is possible, making LDA scalable without losing considerable accuracy. Because of these properties LDA has gained remarkable popularity and many optimized implementations are available. LDA experiments for cross-lingual IR have been published [Cimiano et al., 2009] with disappointing outcomes compared to other techniques. These results do not convincingly show the inappropriateness of LDA, as the authors have embedded multilingual Wikipedia articles without any length normalization or model adaptation. For a method that tries to model characteristics of term statistics, the biggest increase in data-likelihood with limited model capacities should be expected by effectively capturing the predominant language vocabulary instead of doing topical modeling. An indicator that this effect might indeed be at work is that in another run of experiments, trained with in-domain data that have almost constant length ratio (because they are parallel), LDA performs as well as the other methods. Dealing with the supposed length ratio problem could be done either by processing the data accordingly or by adapting the model in a similar fashion as suggested by Ni et al. [2009] for web-page classification (see also [Mimno et al., 2009]). In chapter 3.2.4 we describe in more detail how LDA can be applied in a multilingual setting.

A recently proposed similarity measure that is based on simple ideas and also uses wikipedia as a training corpus is Explicit Semantic Analysis (ESA) [Gabrilovich and Markovitch, 2007, Potthast et al., 2008, Sorg and Cimiano, 2009]). It can be applied in monolingual or cross-lingual settings. The main idea is that terms are represented as weighted vectors of documents in a training corpus. Most often, the column vectors of a tf.idf weighted document-term matrix constructed from the training corpus are used for that. Similarity between words is measured as a distance between such weight vectors. Documents and queries are represented by combining word vectors, e.g. by summing them up. The evaluation results on the method are inconsistent: while it has been found to work well on mate-retrieval tasks in a multilingual setting and is reported to surpass LDA there [Cimiano et al., 2009], in an actual query-based document retrieval setting it showed no competitive performance [Sorg and Cimiano, 2008]. These findings may be explained in different ways: for the first finding there might have been problems in how LDA was applied and compared, for the second finding, no interpolation with a word-vector based representation was undertaken, as it is usually done for dimensionality reduction techniques such as LSA and pLSA [Hofmann, 2001]. In contrast to LDA, which started off from a mathematical formulation, ESA was developed from plausibility considerations and tuned in experimental settings, which might render optimal parametrizations instable in different experiments. Some characteristics make ESA an interesting method to be included in our experiments: Conceptual simplicity, scalability, being a novel approach and being trained on a comparable corpus by default. In chapter 3.3 we discuss monolingual and multilingual ESA.

### 2.2.2 Parallel Corpora

Besides beeing comparable, parallel corpora add the possibility of automatically extracting one-to-one correspondences between phrases and words in texts of different languages. In the following, possibilities of additionally including such more fine-grained information are discussed.

In [Bader and Chew, 2008] the fact is used that SVD of a matrix $X$ can be computed by the Eigenvalue Decomposition of the composite matrix

$$B = \begin{pmatrix} 0 & X \\ X^T & 0 \end{pmatrix}$$

The authors include weighted word–word alignment counts in the upper left block of the matrix (in case of a parallel document, the document vector may contain non-zero entries in term dimensions of more than one langauge). They evaluate their system on the performance of retrieving translations of Bible verses and show a modest but statistically significant improvement on just performing LSA without term alignments. A coherent conceptual interpretation of the overall embedded matrix is hard to give, the weighting scheme is found experimentally.

Cohn and Hofmann [2001] use pLSA to include as an additional source of information links to a document (from another document). The two distributions that the model captures are a conventional term-document model as well as a model of inlinks $c_l$ and documents:

$$P(w_i, d_j) = P(d_j) \sum_{k=1}^{K} P(w_i|z_k) P(z_k|d_j) \tag{1}$$

$$P(c_l, d_j) = P(d_j) \sum_{k=1}^{K} P(c_l|z_k) P(z_k|d_j) \tag{2}$$

The two factorizations are connected via the topic distribution conditioned on a document, $p(Z|d_j)$. The overall likelihood function indicates how well the two factor-izations explain both kinds of (normalized) count data, weighted against each other by a factor $\alpha$. In a similar fashion, word-translation links could be interpolated with word-document counts. However, we see two problems in such an approach: First, interpretability of the model parts would not be fully consistent for the different "ontological" status a word has as a translation from a lexicon or occurring in a document. Secondly, to us it is not evident that the EM-re-estimation equations in Cohn and Hofmann [2001], while looking reasonable at a first glance, really derive from the initial problem statement. From cursory calculations, without making additional non-trivial assumptions, complexity should increase from quadratic to cubic by adding an additional dimension of included information this way.

An interesting alternative based on LDA has been proposed recently by Boyd-Graber and Blei [2009]. Here, multingual pairings of words are extracted from parallel or comparable corpora in an unsupervised way and assigned to a topic. To increase the quality of the pairs, pairing priors derived from e.g. edit-distances or manually edited lexica can be included. In spite of being theoretically interesting, the authors make reservations with respect to scalability issues and evaluate only on a very small training and testing basis with inconclusive results.

## 2.3 Evaluation

Ideally, base-line systems for comparison in our experiments should be conceptually simple, widely used and trained on the same data. Evaluation tasks should be constructed in a way accepted by the research community and provide reproducible results that allow for direct comparison. General competitions like the Cross-Language Evaluation Forum (CLEF, [Peters and Braschler, 2001]) as well as datasets allowing for more specific comparisons are to be taken into consideration.

Two evaluation scenarios are possible and common in the research community. The first are mate retrieval experiments, performed on parallel or comparable document collections with the task of finding the corresponding counterpart of a document in another language. The second scenario consists in classical retrieval settings with queries in one language and relevance-assessed documents in another.

The advantage of a mate-retrieval approach to evaluation is that there are many such corpora available having been used for evaluation, starting with the Bible (as in [Chew and Abdelali, 2007, Bader and Chew, 2008]), Wikipedia (as in [Potthast et al., 2008]), translated news collections and parliament debates (as in [Boyd-Graber and Blei, 2009]), the Official Journal of the European Community and european law texts (as in [Potthast et al., 2008, Cimiano et al., 2009, Sorg and Cimiano, 2009, 2008]). Arguably, such settings also show a certain proximity to document clustering and recommendation (see also [Ni et al., 2009, Wu and Oard, 2008, Olsson et al., 2005]). On the other hand, it is rather an unrealistic setting to appear in real word information retrieval scenarios and a maybe too easy task as well. Also, to see that there is a lot of evaluation done on such corpora does not mean that the results are comparable. In fact, no comparable mate-retrieval results of two different research groups are known to us, unclear preprocessing and selection strategies even complicate the reproduction of experiments and results.

A nicer setup is provided by IR challenges such as CLEF. Here, the evaluation is implemented in clear and replicable ways with hundreds of thousands of documents categorized by a two-stage process of pooling and human relevance-assessment. Comparison can be made immediately with state-of-the-art methods (although such an evaluation may possibly turn out unrewarding for experimental or novel approaches). Also, there are usually only a few, often just about 50 queries, making evaluation unstable to effects like missing vocabulary items.

For our experiments, we evaluate on two datasets: To establish comparability with the findings on LDA and ESA in [Cimiano et al., 2009] and [Sorg and Cimiano, 2009], an evaluation on the Multext JOC corpus is undertaken. To establish comparability to generally applied retrieval methods and give a more realistic picture of performance in real world tasks an evaluation on a CLEF collection is carried out. One of the most accessible of these collections is the CLEF ad-hoc dataset from the year 2000. For the ease of processing and sanity checks, the experiments are done on the language pair German-English.

## 3 Models and Theory

The models described in this chapter are all able to represent documents in vector representations that are not direct bag-of-words vectors of the vocabulary terms, and may differ from them drastically in dimensionality or other qualities. Because these representations can be made directly comparable for more than one language, such models are interesting for cross-language retrieval, especially when the training of them requires less strict properties of the training material than are required for machine translation. For all models introduced in this section comparable corpora are sufficient, while all common statistical machine translation systems need parallel training corpora.

In monolingual retrieval such models are usually justified by the existence two problems: The first is synonymy, the case that two words bear the same meaning while having different surface form, possibly leading to an underestimation of the similarity of two texts. The second is polysemy, possibly leading to overestimation of the similarity of documents in which the same word is used with different meanings.

We include three such models, two probabilistic ones, probabilistic Latent Semantic Analysis and Latent Dirichlet Allocation, and one explicit concept model based on weighting schemes (Explicit Semantic Analysis). Such techniques are often referred to under the term of dimensionality reduction. While for the probabilistic techniques the parameter space in which the documents are represented is in practice indeed of lower dimensionality than the vocabulary, for Explicit Semantic Analysis the new space might be even of bigger dimensionality than the original vocabulary space. Still, the vector representation is smoother than the original one based on words, providing a gradual similarity measure for non-matching terms. In that it is not so different from pseudo-relevance feedback techniques since the connection between terms is established via strongly associated (or highly ranked) documents.

While the models are applicable to any collection of so-called two-mode date (discrete two-dimensional co-occurrence counts), we will consistently use a terminology that immediately draws the connection to modeling of text collections.

### 3.1 Probabilistic Latent Semantic Analysis

Probabilistic Latent Semantic Analysis (pLSA) [Hofmann, 1999, 2001] is a latent variable model that starts with a straightforward model statement and is trained by maximizing the likelihood of the parameters for a training corpus. The most important assumption in pLSA is that the probability distribution of words in a document is only dependent on the distribution of topics in that document. This assumption is the same in more recently developed models.

For several reasons we want to discuss pLSA: It is the first latent variable model widely applied to document clustering and retrieval. Its basic architecture was influential for its successors. It is easily understandable and derivable. It can be easily adapted and implemented. We will discuss to the shortcomings of pLSA in sections 3.1.2 and 3.2.

#### 3.1.1 Probabilistic Model

The pLSA model starts with a collection of $N$ documents over a vocabulary of $M$ words. The document collection is represented by co-occurrence counts $n(d_i, w_j)$, indicating how often document $d_i$ contains word $w_j$. Given such a document collection, the aim of pLSA is to find a probability distribution $P(D, W)$ (of document and word random variables that can take on values $d_i$ and $w_j$ respectively) that maximizes the likelihood of $n(d_{(.)}, w_{(.)})$. While stated in that way $P(D, W)$ would turn out just to be the relative frequencies, the additional assumption is made that $w_i$ and $d_j$ are independent conditioned on the value of a topic variable $Z$ that can take on $K$ different values which are denoted by $z_k$.

From this it follows that the joint probability can be factored, making use of the independence assumption in the last step:

$$\begin{aligned} P(d_i, w_j) &= P(d_i)P(w_j|d_i) \\ &= P(d_i)\sum_{k=1}^{K} P(w_j|z_k, d_i)P(z_k|d_i) \\ &= P(d_i)\sum_{k=1}^{K} P(w_j|z_k)P(z_k|d_i) \end{aligned}$$

This factorization is sometimes called *asymmetric* because the probability distributions involving words and documents are conditioned adversely with respect to the topic variable. The asymmetric factorization can also be explained as a generative process of the three steps:

1. select a document $d_i$
2. select a latent class $z_k|d_i$
3. generate a word $w_i|z_k$

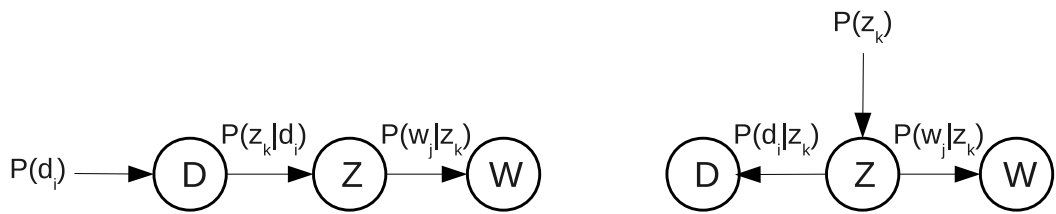

Figure 1: PLSA model in graphical notation. **Left:** Asymmetric parametrization. **Right:** Symmetric parametrization.

One might call this process not fully generative since a document is picked from a preexisting set and not generated by a random process. The model is determined by $KN + KM$ parameters, which is (assuming $k \ll M, N$) considerably less than $MN$ as in the case without the conditional independence assumption. The variable $Z$ functions as bottleneck establishing the interdependence between documents and words. From the equivalent *symmetric* factorization given in equation 4 some parallels to Latent Semantic Analysis [Deerwester et al., 1990], based on truncated singular value decomposition, can be drawn: the values $P(d_i|\cdot)$ and $P(w_j|\cdot)$ play a role similar to that of the left and right singular vectors: in both cases co-occurrence matrices can be constructed from them for each of the assumed topics; $P(z_k)$ corresponds to singular values by weighting the topical co-occurrence matrices.

$$P(d_i, w_j) = \sum_{k=1}^{K} P(d_i, w_j | z_k) P(z_k) \qquad (3)$$

$$= \sum_{k=1}^{K} P(d_i|z_k) P(w_j|z_k) P(z_k) \qquad (4)$$

When the model is stated, the next step is to find a suitable parametrization, that is values for all $p(d_i)$, $P(w_j|z_k)$ and $P(z_k|d_i)$. Often one works with the asymmetric factorization because it characterizes the documents by their topic distribution. Parameters are estimated according to the maximum likelihood principle, so that those parameters are sought which give rise to the highest probability of the observed data. The likelihood function becomes:

$$\mathcal{L} = \prod_{i=1,j=1}^{N,M} P(d_i, w_j)^{n(d_i,w_j)}$$

For ease of calculation, optimization is done on the log of this function, which gives

$$\begin{aligned}
\mathcal{L}^* &= \sum_{i=1,j=1}^{N,M} n(d_i, w_j) \ \log p(d_i, w_j) \\
&= \sum_{i=1}^{N,M} n(d_i) \ \log P(d_i) + \sum_{i=1,j=1}^{N,M} n(d_i, w_j) \ \log \sum_{k=1}^{K} P(w_j|z_k) P(z_k|d_i)
\end{aligned}$$

where $n(d_i) = \sum_{j=1}^{M} n(d_i, w_j)$ is the document length. Taking the derivative with respect to $P(d_i)$ and setting to zero yields $P(d_i) \propto n(d_i)$. The other parameters cannot be determined analytically because they involve the variable $Z$ for which no values can be observed from the data. For problems of this type the expectation-maximization (EM) algorithm offers a solution (see page 16 for a general introduction). In our case the steps of the algorithm become:

1. **E-step:** Calculate $P(Z|D, W)$.

2. **M-step:** Find new $P(W|Z)$ and $P(Z|D)$ - as mentioned before, $P(D) \propto n(D)$ can be estimated independently.

The expected complete data likelihood for pLSA is given by

$$E\left[\mathcal{L}^c\right] = \text{const} \ + \sum_{i=1,j=1,k=1}^{N,M,K} P(z_k|d_i, w_j) \ n(d_i, w_j) \log \left[P(w_j|z_k) P(z_k|d_i)\right]$$

where const is a quantity only dependent on observable counts. For standard pLSA, we refer to [Hofmann, 2001] for a full derivation. See section 3.1.4 for our multilingual adaptation of this model.

The estimate of the hidden variable, given the values of the observed variables, can be **The EM Algorithm** [Dempster et al., 1977, Bilmes, 1998] aims at finding a maximum-likelihood solution for problems involving observable and non-observable (hidden) variables. Denote by $X$ values the observable data have taken on and by $Z$ one possible assignmet of values to the hidden variables (we discuss discrete variables only). A parametrization of a joint probability disbution over all variables is denoted by $\Theta$.

We denote by $P(X, Z|\Theta)$ the *complete data likelihood*, a function of $\Theta$ for a fixed assignment to both observed and unobserved variables. The straighforward optimization goal with respect to the observed data, obtained by marginalizing over all possible assignments to the hidden variables

$$\hat{\Theta} = \arg\max_{\Theta} \log \left\{ \sum_{Z} P(X, Z|\Theta) \right\} \qquad (5)$$

is often analytically intractable. EM is motivated by the observation that the *complete data log-likelihood*

$$\mathcal{L}^c = \log P(X, Z|\Theta)$$

is often easy to optimize. However, the assignment of $Z$ is unknown. Therefore, the expectation of this value with respect to $Z$ is used, while $Z$ is taken to be distributed according to a previous parameter estimate $\Theta^{i-1}$.

The key finding for EM is that the iterative optimization for $\Theta^i$ of the expected complete-data log-likelihood

$$E[\mathcal{L}^c|X, \Theta^{i-1}] = \sum_{Z} P(Z|\Theta^{i-1}) \log P(X, Z|\Theta^i) \qquad (6)$$

never decreases the observable data-likelihood [Dempster et al., 1977, Wu, 1983]. Often $E[\mathcal{L}^c]$ or $Q(\Theta^i, \Theta^{i-1})$ is written short for $E[\mathcal{L}^c|X, \Theta^{i-1}]$. Note, that in formula (6) the summation appears outside the logarithm which makes it analytically more appealing than formula (5).

The algorithm can then be split up into two alternating steps:

1. **E-step:** Estimate how often the hidden variables take on certain values. Use the last parameters $\Theta^{i-1}$.

2. **M-step:** Find new parameters $\Theta^i$ that maximize the expected log-likelihood of both observed and hidden variables, using the estimates of the last step.

In practice, the parameters are often initialized with random values.

read off immediately from the original problem statement, making use of the independence assumption and Bayes' formula:

$$P(z_k|d_i, w_j) = \frac{P(w_j|z_k)P(z_k|d_i)}{\sum_{k=1}^{K} P(w_j|z_k)P(z_k|d_i, w_j)}$$

The optimal parameters are found by taking the derivative of the complete data likelihood function $E[\mathcal{L}^c]$ with respect to each parameter and setting to zero. Because the maximizing parameters have to form probability distributions, appropriate Lagrange multipliers have to be added for normalization. The resulting re-estimation equations are:

$$p(w_j|z_k) = \frac{\sum_{i=1}^{N} n(d_i, w_j)P(z_k|d_i, w_j)}{\sum_{i=1,m=1}^{N,M} n(d_i, w_m)P(z_k|d_i, w_m)}$$

$$P(z_k|d_i) = \frac{\sum_{j=1}^{M} n(d_i, w_j)P(z_k|d_i, w_j)}{n(d_i)}$$

### 3.1.2 Theoretical Considerations

Let us call a parametrization $\Theta$ (it comprises all the values found for $p(d_i)$, $P(w_j|z_k)$ and $P(z_k|d_i)$) and the data $D$ (it is the statistics for the documents in our training corpus). For the maximium likelihood estimate, we look for a $\hat{\Theta}$ maximizing $P(D|\Theta)$. This is called the maximum a posteriori estimate with an uninformative (uniform) prior $p(\Theta)$ in Bayesian statistics:

$$\hat{\Theta} = \arg\max_{\Theta} p(D|\Theta)p(\Theta)$$

Using this estimate has several disadvantages. The main criticism from a Bayesian point of view is that a particular choice of $\hat{\Theta}$ might maximize the distribution over the parameters $p(\Theta|D)$ but does not account for our uncertainty regarding this choice[2] and entirely excludes other parametrizations that might be similarly reasonable but happen to have received a little less evidence from the data. The more practical argument is therefore that maximum likelihood estimates tend to overfit. In a fully Bayesian approach the probability of the data should account for different possible parametrizations and also include a non-uniform prior distribution $p(\Theta|\lambda)$ over the model parameters, where all the intuitions about the problem are captured in a set of hyper-parameters $\lambda$.

---

[2] it holds that $p(D, \Theta) \propto \frac{p(D,\Theta)}{p(D)} = p(\Theta|D)$, since $p(D)$ is unaffected by different $\Theta$

We want to contrast the maximum likelihood (formula 7) and the Bayesian (formula 9) approach, both assigning a probability estimate to some new data items $D_{new}$, given training data $D_{old}$. In both cases the assumption is made that the probability of the data is only dependent on the parametrization.

$$
\begin{aligned}
p(D_{new}|D_{old}) &= P(D_{new}|\arg\max_{\Theta} p(D_{old}|\Theta)) & (7) \\
&= P(D_{new}|\hat{\Theta}) & (8)
\end{aligned}
$$

$$
\begin{aligned}
p(D_{new}|D_{old},\lambda) &= \int P(D_{new}|D_{old},\Theta,\lambda)d\Theta & (9) \\
&= \int P(D_{new}|\Theta)p(\Theta|D_{old},\lambda)d\Theta & (10)
\end{aligned}
$$

Many problems naturally disappear when a Bayesian approach is undertaken, for example the problem that with pLSA $p(d)$ distributes its probability mass only about documents present in the training corpus. Other problems come up, for example how to find a suitable $\lambda$ and how to find a solution for the parameters in a tractable way. In chapter 3.2 we will discuss a model that quite successfully offers a solution to most of these problems.

### 3.1.3 Practical Issues

In the EM-algorithm the following equations have to be repeatedly evaluated:

$$
\begin{aligned}
p(z_k|d_i,w_j) &= \frac{p(w_j|z_k)p(z_k|d_i)}{\sum_{k=1}^{K} p(w_j|z_k)p(z_k|d_i)} \\
p(w_j|z_k) &= \frac{\sum_{i=1}^{N} n(d_i,w_j)p(z_k|d_i,w_j)}{\sum_{i=1,m=1}^{N,M} n(d_i,w_m)p(z_k|d_i,w_m)} \\
p(z_k|d_i) &= \frac{\sum_{j=1}^{M} n(d_i,w_j)p(z_k|d_i,w_j)}{n(d_i)}
\end{aligned}
$$

In a naive implementation this would lead to space and time requirements of $O(NMK)$ because of the first of the three formulas. However, because $p(z_k|d_i,w_j)$ is multiplied only with corresponding counts $n(d_i,w_j)$ only values have to be evaluated where these counts are non-zero. The complexity is therefore only $O(fK)$, with $f = \sum_{i=1,m=1}^{N,M} \mathbf{1}_{n(d_i,w_m)>0}$. We have adapted matlabs vector multiplication for this end so that it evaluates the outcome only at specified positions of the result matrix.

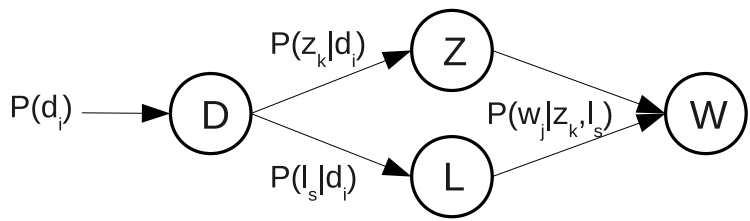

Figure 2: The multilingual pLSA model in graphical notation. Word probabilities are only dependent on the topic and language distribution of a document.

The pLSA model has been reported to get stuck in local optima and to overfit. Two strategies have been applied to alleviate these issues. The first is tempering, an annealing method that tries to increase the entropy of the posterior distribution $p(Z, D, W)$ by taking it to the power of a decreasing $0 < \beta \leq 1$ in the re-estimation equations (details can be found in [Hofmann et al., 1999, Ueda and Nakano, 1998]). Another often applied technique is to interpolate the parameters of different runs. For information retrieval, the pLSA model is usually interpolated with a weighted word-vector model.

### 3.1.4 Multilingual pLSA

In this section we introduce our own pLSA model that is designed to capture the topical composition of multilingual documents. We want to assume a document that consists of parts in different languages which need not be equally long. For every word it is observable to which language part it belongs. In order to model multilinguality we introduce an additional variable $L$ that can take on $S$ different values indicating a particular language. We now have two bottleneck variables, $Z$ and $L$, of which one is observable, the other not. We carry over the independence assumptions made for $Z$ to the case with two such variables: The language and the topic that generate a word are independent conditioned on the document. Words and documents are independent conditioned on language and topic. The generative process is then:

1. select a document $d_i$

2. select a latent class $z_k|d_i$

3. select a language $l_s|d_i$

4. generate a word $w_j|z_k, l_s$

We still have only one latent variable $Z$ since we can observe the language variable. The joint probability of a word $w_j$ occurring in a document $d_i$ can be written as:

$$\begin{aligned} p(d_i, w_j) &= p(d_i)p(w_j|d_i) \\ &= p(d_i) \sum_{s=1}^{S} \sum_{k=1}^{K} p(w_j|z_k, l_s, d_i) p(z_k, l_s|d_i) \end{aligned}$$

Using the independence assumption and the fact that we can observe the language of a word, that means $p(w_j|z_k, l_s) = 0$ for all languages $l_s$ other than the language $s(w_j)$ of the word considered, we get:

$$p(d_i, w_j) = p(d_i, s(w_j)) \sum_{k=1}^{K} p(w_j|z_k, s(w_j)) p(z_k|d_i)$$

From Bayes' law and using the knowledge about $p(w_j|z_k, l_s)$ as above, we get:

$$p(z_k|d_i, w_j) = \frac{p(w_j|s(w_j), z_k) p(z_k|d_i)}{\sum_{k'} p(w_j|s(w_j), z_{k'}) p(z_{k'}|d_i)}$$

Similarly to the monolingual case one immediately gets $p(d_i, l_s) \propto \sum_{\{w_j|s(w_j)=l_s\}} n(d_i, w_j)$ by taking the derivative and setting to zero. For the parameters involving the latent variable one derives EM with the complete data likelihood (omitting the constant dependent on observable counts):

$$E\left[\mathcal{L}^c\right] = \sum_{i=1, j=1, k=1}^{N, M, K} n(d_i, w_j) p(z_k|d_i, w_j) \log\left[p(w_j|z_k, s(w_j)) p(z_k|d_i)\right]$$

Adding Lagrange multipliers to ensure that probabilities add up to one:

$$\begin{aligned}
\mathcal{H} &= \sum_{i=1,j=1}^{N,M} n(d_i, w_j) \sum_{k=1}^{K} p(z_k|d_i, w_j) \log\ p(w_j|z_k, s(w_j)) \\
&+ \sum_{i=1,j=1}^{N,M} n(d_i, w_j) \sum_{k=1}^{K} p(z_k|d_i, w_j) \log\ p(z_k|d_i) \\
&+ \sum_{k=1,s=1}^{K,S} \tau_{k,s}(1 - \sum_{\{w_j|s(w_j)=l_s\}} p(w_j|z_k, s(w_j))) \\
&+ \sum_{i=1}^{N} \rho_i(1 - \sum_{z_k} p(z_k|d_i))
\end{aligned}$$

After taking the derivative with respect to $p(w_j|z_k, s(w_j))$ and setting to zero, one gets:

$$p(w_j|z_k, s(w_j)) = \frac{\sum_{i=1}^{N} n(d_i, w_j) p(z_k|d_i, w_j)}{\tau_{k,s(w_j)}}$$

Inserting this for $p(w_j|z_k, s(w_j))$ we can optimize for the Lagrange multipliers $\tau_{k,l}$, and get:

$$\begin{aligned}
\frac{\delta}{\delta \tau_{k,l}} \mathcal{H} &= \frac{\delta}{\delta \tau_{k,l}} \sum_{i,j} n(d_i, w_j) \sum_{k} p(z_k|d_i, w_j) \ln \frac{\sum_i n(d_i, w_j) p(z_k|d_i, w_j)}{\tau_{l,s(w_j)}} \\
&+ \cdots + \sum_{l,k} \tau_{l,k} \left(1 - \sum_{\{w_j|s(w_j)=l_s\}} \frac{\sum_i n(d_i, w_j) p(z_k|d_i, w_j)}{\tau_{l,k}}\right) \\
&= \frac{\delta}{\delta \tau_{l,k}} \tau_{l,k} - \sum_{i,\{w_j|s(w_j)=l_s\}} n(d_i, w_j) p(z_k|d_i, w_j) \ln \tau_{l,k} \\
&= 1 - \frac{\sum_{i,\{w_j|s(w_j)=l_s\}} n(d_i, w_j) p(z_k|d_i, w_j)}{\tau_{l,k}} = 0
\end{aligned}$$

In the first step, we omitted all summands that do not contain $\tau_{k,l}$. Finally the values for $\tau_{k,l}$ can be reinserted. An analogous calculation for $\rho_i$ gives us all necessary re-estimation equations:

$$p(w_j|z_k, s(w_j)) = \frac{\sum_{i=1}^{N} n(d_i, w_j) p(z_k|d_i, w_j)}{\sum_{i,\{w_j|s(w_j)=l_s\}}^{N} n(d_i, w_j) p(z_k|d_i, w_j)}$$

$$p(z_k|d_i) = \frac{\sum_j n(d_i, w_j) p(z_k|d_i, w_j)}{n(d_i)}$$

Our model can work on documents in any number of different languages, with arbitrary language proportions, including monolingual documents in a multilingual setting. Strict separation of content and language variables avoid estimation of the language by the topic variables, so that a unified comparison of the semantic content of multilingual documents is possible. In section 4.1 we apply this model in a retrieval task on multilingual Wikipedia documents.

### 3.2 Latent Dirichlet Allocation

#### 3.2.1 Probabilistic Model

Latent Dirichlet Allocation (LDA) [Blei et al., 2003, Griffiths and Steyvers, 2004, Steyvers and Griffiths, 2007] is a latent variable model that gives a fully generative account for documents in a training corpus and for unseen documents. Each document is characterized by a topic distribution, words are emitted according to an emission probability dependent on a topic. The main difference to pLSA is that both topic distributions and word emission distributions are assumed to be generated by Dirichlet priors. A Dirichlet distribution with $T$ parameters $Dir(\alpha_1 \cdots \alpha_T)$ (short $Dir(\alpha)$) assigns a probability to a $T$-dimensional multinomial distribution $Mult(p_1 \cdots p_T)$ giving the density:

$$\frac{\Gamma(\sum_j \alpha_j)}{\prod_j \Gamma(\alpha_j)} \prod_{j=1}^{T} p_j^{\alpha_j - 1}$$

where $\Gamma$ is the gamma function. The parameters of the Dirichlet distribution determine two properties of the multinomials drawn from it: first, the expected values $p_i$ are proportional to the paramters $\alpha_i$; secondly, the expected variances of the $p_i$ decrease when the sum of the $\alpha$'s gets bigger. When using the Dirichlet as a prior, it is common to assign the same value $\alpha = \alpha_1 = \cdots = \alpha_T$ to all parameters and basically to direct only the "peakiness" of the multinomials drawn from it: values for $\alpha$ bigger than 1 favor uniform multinomials (the mode of the Dirichlet is for a uniform multinomial), values smaller than 1 punish uniform multinomials (the Dirichlet has modes at multinomials that assign all probability mass to one event).

The LDA model describes the process of generating text in the following way:

1. For all $k$ topics generate multinomial distributions $\psi^{(z_k)} = p(w_j|z_k) \sim Dir(\beta)$.

2. For every document $d$:

    a) Generate a multinomial distribution $\theta^{(d)} = p(z_k|d) \sim Dir(\alpha)$.

    b) Generate a document length, and topics $z_i \sim \theta^{(d)}$ for every position $i$ in the document.

    c) Generate words $w_i \sim \psi^{(z_i)}$ for every position in the document.

Usually, no generative account for the length of the document is given, because it has no impact on the other parts of the model one is interested in. Figure 3 shows the generative model in plate notation.

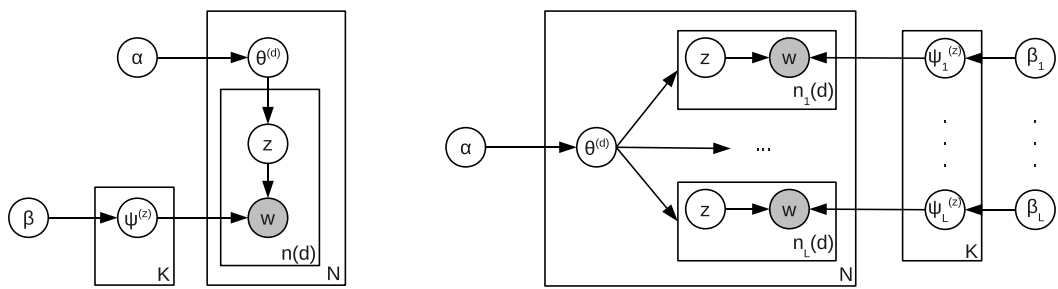

Figure 3: Latent Dirichlet Allocation in plate notation. **Left:** Standard (monolingual) model. **Right:** Multilingual model.

### 3.2.2 Sampling for LDA

The first approach to estimate such a model [Blei et al., 2003] was to represent and estimate $\psi$ and $\theta$ explicitly, resulting in different inference tasks to be solved and combined. Later approaches concentrate on getting a sample of the assignment of words to topics instead [Griffiths and Steyvers, 2004]. We now give a short survey of the conceptual steps involved in the topic sampling process. For a more detailed description of the theory behind sampling, we refer to general text books such as [Newman and Barkema, 1999, MacKay, 2003, Bishop, 2006].

Gibbs sampling is a simple technique based on the fact that when the outcome of one variable is sampled conditioned on the previously sampled outcomes of the other variables, the overall distribution converges, when sampled iteratively, to the underlying distribution.

For LDA, this sampling is particularly easy because the distributions involved are multinomials conditioned on Dirichlet priors. The Dirichlet distribution is conjugate to the multinomial distribution. This means that when the inital assumption about a multinomial is that it is drawn from a Dirichlet (*prior*) distribution with parameters $\boldsymbol{\alpha}$, then, after observations $\mathbf{x}^{(i)}$ (vectors indicating the outcome) generated by the multinimial are available, the most likely underlying (*posterior*) distribution for having generated the multinomial is again a Dirichlet distribution with parameters

$$\boldsymbol{\alpha}' = \boldsymbol{\alpha} + \sum_i \mathbf{x}^{(i)}$$

One might want to regard the initial parameters $\boldsymbol{\alpha}$ as pseudo-counts, added before any actual data have been seen.

The expectation of a multinomial distribution $\mathbf{p} \sim \text{Dir}(\boldsymbol{\alpha})$ is

$$E[\mathbf{p}_i] = \frac{\boldsymbol{\alpha}_i}{\sum_j \boldsymbol{\alpha}_j}$$

Therefore, the estimates of the word emission probabilities $\psi^{(z_k)} = p(w_j|z_k)$ and of the topic probability $\theta^{(d)} = p(z_k|d)$, conditioned on the *posterior* estimate, as observations including samples for all positions, are:

$$\hat{\psi}_j^{(w)} = \frac{n_j^{(w)} + \beta}{n_j^{(\cdot)} + W\beta}$$

$$\hat{\theta}_j^{(d)} = \frac{n_j^{(d)} + \alpha}{n_\cdot^{(d)} + T\alpha}$$

The sampling equation can easily be obtained by excluding the current position from the observations:

$$P(z_i = j|\mathbf{z_{-i}}, \mathbf{w}) = \frac{P(z_i = j, \mathbf{w}|\mathbf{z_{-i}})}{P(\mathbf{w})}$$

$$= \hat{\psi}_{-i,j}^{(w_i)} \hat{\theta}_{-i,j}^{(d_i)} \frac{P(\mathbf{w_{-i}}|\mathbf{z_{-i}})}{P(\mathbf{w})}$$

$$\propto \frac{n_{-i,j}^{(w_i)} + \beta}{n_{-i,j}^{(\cdot)} + W\beta} \frac{n_{-i,j}^{(d_i)} + \alpha}{n_{-i,\cdot}^{(d_i)} + T\alpha}$$

Here, $n_{-i,j}$ counts how often a topic $j$ (using the dot $\cdot$ for any topic) has been assigned to any position in the text collection, excluding the position $i$, for which a new topic is being sampled. The superscripts denote restrictions on the counts, for example $(w_i)$ only includes assignments to positions that have the same word assigned as at position $i$ and $(d_i)$ only includes assignments to positions that belong to the same document as position $i$. The dot $(\cdot)$ indicates no restriction.

The sampling distribution can be interpreted the following way: The left formula, corresponding to a smooth estimate of $p(w_i|z_j)$, favors topics that have often generated the same word as at position $i$. The right formula, providing a smooth estimate of $p(z_j|d)$, favors topics that have already often occurred in the current document.

Sampling itself does not distinguish between training and inference documents. In practice, the topic counts are fixed for the training set after sampling is assumed to have reached a stable state. Sampling iterations for inference change the topic assignments only for single unseen documents.

### 3.2.3 Practical Issues

To determine the similarity between two documents, one can compare either their sampled topic vectors or the probability vectors obtained from them [Griffiths and Steyvers, 2004]. When other variational EM estimation techniques are applied, also other parameter vectors might be available and used [Cimiano et al., 2009, Blei et al., 2003]. The comparison between these vectors can be done either by taking the cosine similarity of their angles or, in case the vector is indeed a probability distribution (as for $\theta$), by using probability divergence measures, such as the (symmetrized) Kullback Leibler divergence or the Jensen-Shannon divergence.

For language model based information retrieval, one is interested in the probability of a query, given a document $p(q|d_i)$. Wei and Croft [2006] interpolate a language model based on LDA with a unigram language model directly estimated on the document. The LDA model gives smooth estimates for every word in the query by $\sum_j P(w|z=j)P(z=j|d_i)$. With this formulation, no latent topics are estimated for the query and therefore no $\theta$ exists that could disambiguate terms in the query based on other terms occurring in it (in contrast to topics having been sampled following a combination of word-based $\psi$ and query-based $\theta$).

LDA can be easily parallelized [Newman et al., 2007, Wang et al., 2009]. Usually, the set of training documents is split and sampling is done independently in each iteration. However, the topic-per-word counts have to be updated for all processes in each iteration. Therefore, the speedup is linear for the size of the corpus, but no speedup is obtained for the vocabulary size. For large vocabularies and few documents communication costs may outweigh any speedup obtained.

Whenever a sampling technique is used, one wants to be sure that the estimates are stable. This could be a problem when only one topic is sampled per position for short documents or queries. The most natural way to overcome this problem is to average the results of several sampling iterations.

### 3.2.4 Multilingual LDA

LDA has been extended for several languages [Ni et al., 2009], see also [Mimno et al., 2009] for an investigation of the semantic clustering of this model. Most of the components remain the same, the main difference is that for each language $l_s$ a different word emission distribution $\psi_{l_s}$ is assumed. Depending on the language of a position in a document, a word is generated conditioned on the corresponding distribution. The sampling equation becomes, using similar notation as before:

$$P(z_i = j|\mathbf{z_{-i}}, \mathbf{w}) \propto \frac{n_{-i,j}^{(w_i,l_i)} + \beta_{l_i}}{n_{-i,j}^{(l_i)} + W_{l_i}\beta_{l_i}} \frac{n_{-i,j}^{(d_i)} + \alpha}{n_{-i,\cdot}^{(d_i)} + T\alpha}$$

Figure 3 shows the model in plate notation. This model allows to deal with multilingual topics in an elegant way. The model is truly multilingual in that it does not use the topic variable to estimate the language, as it could be the case for a monolingual model applied to a multilingual document without any adaptions on either the data or the model side. A theoretically sound model does not mean that it also provides a good bridging between two languages. It is crucial [Mimno et al., 2009] how many "glue documents", i.e. documents that indeed have counterparts in all compared languages, are available. Although the topic variables may not estimate the language, the partition of the topic space might diverge. Consider, as an extreme example, a multilingual model trained on hundreds of thousands of documents, each available in exactly one of the covered languages, and only one document providing text in all of them. It becomes obvious that this is clearly too little to align the topic spaces between the languages.In our weighting experiments in section 4.1 we show evidence that this intuition can be generalized: not only should a large number of glue documents exist, good bridging documents should optimally be of equal length. We believe, therefore, that it is a more promising approach to normalize the data suitably to fit into a standard LDA model. Moreover, this has the advantage that one of the many optimized toolkits readily available for LDA can be used out of the box, without the need to adapt their sampling schemes.

We give an informal argument why a standard (monolingual) LDA model trained on a normalized multilingual document collection, with all language parts in a document being of equal length, is essentially equivalent to the multilingual LDA model applied to the same collection. Consider $L$ languages, and the same prior parameters $\beta$ in all languages. We want to denote with

$$\zeta_{i,j} = \frac{n_{-i,j}^{(l_i)} L}{n_{-i,j}^{(\cdot)}}$$

the proportion of topic occurrences in a language to occurrence counts in all languages. We assume that the words can be unequivocally identified as belonging to a particular language (this is in practice easily achieved by attaching suitable prefixes) so that $n^{(w_i, l_i)} = n^{(w_i)}$. For simplicity, we also assume same vocabulary size in all languages (without this assumption smoothing differs per language). We start with the multilingual model:

$$P(z_i = j | \mathbf{z_{-i}}, \mathbf{w}) \quad \propto \quad \frac{n_{-i,j}^{(w_i)} + \beta}{\frac{\zeta_{i,j}}{L} n_{-i,j}^{(\cdot)} + \frac{1}{L} W \beta} \frac{n_{-i,j}^{(d_i)} + \alpha}{n_{-i,\cdot}^{(d_i)} + T\alpha}$$

For $\zeta_{i,j} = 1$ it samples just as the monolingual model since the constant factor $L$ can be ignored. For $\zeta_{i,j} < 1$ the denominator is decreased, making the topic (that is less frequent for the language at position $i$ than for others) more likely to be sampled for this language. Likewise $\zeta_{i,j} > 1$ makes a more frequent topic less probable in the next iteration. One can argue that in the multilingual model on normalized data a stable situation occurs when the topic distribution of one language equals that of all languages. In this situation the multilingual model behaves like the standard model. The standard model can therefore be used on such data.

## 3.3 Explicit Semantic Analysis

Explicit Semantic Analysis (ESA) [Gabrilovich and Markovitch, 2007, Potthast et al., 2008, Sorg and Cimiano, 2008, 2009] is another scheme to overcome the word-mismatch problem. In ESA, the association strength of words to the documents in a training collection is computed, and vectors of these asscociations are used to represent the words in the semantic space spanned by the document names. These word representations can be used to compare words, they can also be combined for a comparison of texts.

### 3.3.1 Formalization

Several formalizations are possible in this setting. The fundamental ingredients that determine an implementation are:

- **Word vectors:** For every word $w$ a vector $\mathbf{w}$, indicating its association strength to the documents is computed. Formally:

$$\mathbf{w} = \langle as(w, a_1), \cdots, as(w, a_N) \rangle$$

  Where $as(w, a_n)$ indicates the association strength of $w$ to training document $a_n$ (mnemonic for **a**rticle, since the most commonly used articles are Wikipedia documents) of $N$ training documents in total.

- **Text vectors:** For a new document (or query) $d$ a representation $\mathbf{d}$ is computed from the word vectors:

$$\mathbf{d} = f(\{\mathbf{w} | w \in d\}_b)$$

  The subscript $b$ should indicate that in the general case one wants to be able to consider the multiset ("bag") of word vectors, allowing for counts of the words in $d$ to be considered.

- **Similarity function:** Giving two documents $d_1$ and $d_2$ the similarity is computed using a similarity function on their text vectors.

The word vectors that were used in [Gabrilovich and Markovitch, 2007] and found to be optimal in [Sorg and Cimiano, 2009] are obtained by taking the respective columns of the tf.idf-weighted document-term matrix $A$ of the training collection. In other words, this corresponds to an association strength function where:

$$as(w, a_n) = \frac{A_{n,w}}{\sum_{w' \in W} A_{n,w'}} \log \frac{N}{\sum_{n'=1}^{N} \mathbf{1}_{A_{n',w}>0}} \tag{11}$$

Here, $\mathbf{1}_{A_{n',w}>0}$ is an indicator that equals to one if word $w$ has appeared in document $a_{n'}$ at least once, and that equals to zero otherwise. Notice that in this formulation the relative term frequency is used. In effect, this is a length normalization, making all documents contribute equally strongly. We use this choice of word vectors in our experiments.

For the text similarity, several settings have been proposed. In [Gabrilovich and Markovitch, 2007] a weigting of the word vectors is used, they are multiplied with scalars equal to, again, an tf.idf weighting of the terms, and then summed up. It is however not very clearly described which exact instantiation of the tf.idf function was used in the experiments. Sorg and Cimiano [2009] explore further combination schemes, including the sum of the elements of either the multiset (considering term frequency) or of the set (not considering term frequency) of word vectors. They find that the set combination works best, yielding preliminary text vectors of the form:

$$\hat{\mathbf{d}} = \sum_{w \in d} \mathbf{w}$$

It is beneficial to truncate the resulting vectors at a certain threshold. The thresholding that turned out to be most successful was to retain the $c$ biggest non-zero values of the vectors $\hat{\mathbf{d}}$. Let $t$ be the value of the $(c+1)$th biggest component of $\hat{\mathbf{d}}$.

$$\mathbf{d_i} = \begin{cases} \hat{\mathbf{d}}_\mathbf{i} & \text{if } \hat{\mathbf{d}}_\mathbf{i} > t, \\ 0 & \text{otherwise.} \end{cases}$$

Here, $\mathbf{d_i}$ refers to the $i$th component of $\mathbf{d}$. In Sorg and Cimiano [2009] it was found that a cut-off value of $c = 10000$ works best. Again, we use this parametrization in all following experiments that involve ESA. As a similarity function the cosine is suggested and used by us.

### 3.3.2 Multilingual ESA

The application of this model in a multilingual setting is straightforward. For $L$ languages consider document term matrices $A^{(1)} \cdots A^{(L)}$. Construct the matrices in a way that the document rows correspond. For all languages each of the rows $A_n^{(\cdot)}$ contains documents about the same topic across the languages. Therefore only documents can be included that are available in all of the considered languages. For each document the mapping to text vectors is performed using a monolingual matrix corresponding to its language. As the documents are aligned, similarities can be computed across languages. Because the relative frequency is used in the tf.idf-weighting, all documents are normalized and no bias occurs for documents longer in one language than in another.

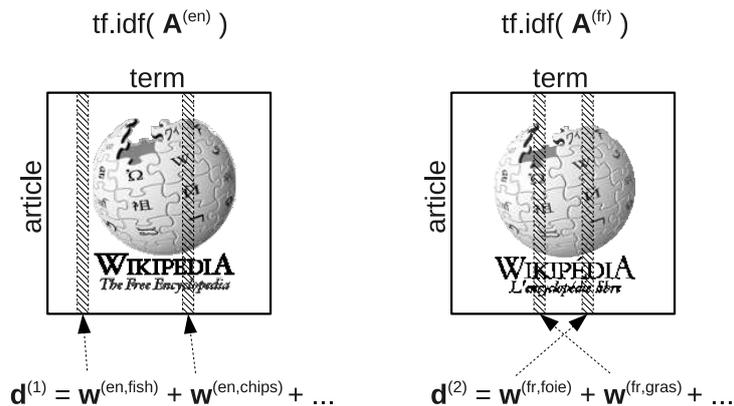

Figure 4: Explicit Semantic Analysis. The example illustrates the construction of an English and a French text vector for food-related documents.

Different comparable text collections can be used. The most easily available and best performing one is Wikipedia, originally ESA was also tried on the Open Dictionary Project (ODP)[3] data, with considerably worse results. Sorg and Cimiano [2008] observe that even when the ESA model is used to compare documents taken from two languages out of the set French, German and English it slightly improves retrieval performance to restrict the training corpus to documents available in all of the three languages. This finding is presented as a side-note rather than substantiated with further evaluation results. Since our work aims at finding best practices for cross-langual retrieval, the most general setting for us is to use training corpora in exactly those languages in which we want to do retrieval.

### 3.3.3 Implementation

Wikipedia dumps can be bigger than the working memory current customary computers possess. The English Wikipedia dump for example comprises 27G of xml. From the dump a sparse tf.idf-weighted document-term matrix is constructed, in a line by line process. This matrix has to be inverted to obtain the word vectors, while considering all documents per word. In the only open academic implementation known to us[4], this invertation is done by using an indexer and a database. In our implementation we use an iterative blockwise invertation which is adaptive to different memory equipments and is usable without installing any external libraries. As our implementation has also put focus on efficient xml processing, we observe a speedup of more than factor 10 when processing the English Wikipedia dump as compared with *research-esa*.

---

[3] http://www.dmoz.org

---

[4] *research-esa*, `http://code.google.com/p/research-esa/`

# 4 Experiments

| Corpus | Retrieval type | Query Set | Target Set | Training Basis |
|:---:|:---:|:---:|:---:|:---:|
| Wiki-Subset | thematical correspondence | $1K$ articles | $1K$ articles | $1-2K$ articles |
| Multext JOC | translations | $3K$ texts | $3K$ translations | $320K$ articles |
| CLEF2000 | query based text retrieval | 33 queries | $110K$ news texts | $320K$ articles |

Table 1: Corpora used for retrieval experiments.

As Table 1 shows, three corpora were used for our experiments: A Wikipedia subset assembled by us, the Multext JOC corpus and the German-English dataset of the CLEF 2000 ad-hoc track. A detailed discussion of the characteristics is given in the respective chapters.

## 4.1 Preliminary Experiments

In the first series of experiments the models are explored on a small dataset. All three models, ESA, LDA and our adaptation of pLSA are compared. As for the latent variable models, it is of particular interest how data weighting and normalization affects retrieval performance. We will fix some decisions and will consider a smaller set of hypotheses in later large scale experiments.

The starting point of this chapter are hypotheses and questions that come up with the theoretical models we introduced and with experiments reported elsewhere. We will summarize them as follows:

1. How does the multilingual pLSA model perform?

2. How important are "glue documents" in this model (that is, truly multilingual documents)? What weighting schemes favor important documents?

3. How does a monolingual LDA model perform on multilingual data
    - in the form reported in [Cimiano et al., 2009]?
    - when using a scheme that takes theoretical considerations into account?

4. How do the latent variable models compare to ESA?

5. How do combinations of them perform?

Figure 5: Document–term matrix of a subset of Wikipedia documents used in the initial experiments.

Weighting scheme experiments are conducted with the pLSA model, because it can handle real valued quantities which can be scaled up continuously. For the LDA model integer valued quantities are necessary because of the sampling process. In any case, weighting schemes are often ad-hoc. For that reason we want to use them only as indicators of important properties and as a justification for restricting our data in a sensible way.

### 4.1.1 Experiment Setup

Our experiments were conducted using a bilingual dataset. We performed a *mate retrieval*-task, that means finding a document in a foreign language that corresponds to a document in the original language. For example in a collection of English Wikipedia articles, find the one dealing with "french fries", given the German article corresponding to "Pommes Frites".

Our dataset was constructed from a German and an English Wikipedia dump[5]. We randomly chose 2000 documents belonging to the German Wikipedia category "Philosophie" for each of which an English Wikipedia article exists to which (1) the German article links and (2) which itself links to exactly this German article. We ignored certain Wikipedia pages that we do not expect to contain encyclopedic knowledge, like redirects or disambiguation pages. Tokenization was done by regarding all characters that are not in a Unicode letter block as delimiters, and all tokens that were only one letter long or were found in the Snowball stopword list were disregarded. We also removed all words that occurred only once in this collection. The remaining words were stemmed using the Snowball[6] stemmer.

---

[5] we used the German snapshot of 2009-07-10 and the English snapshot of 2009-07-13

The data are organized in a matrix of three blocks (see also Figure 5): One *comparable* block and two *monolingual* blocks. The comparable block comprises 1000 multilingual documents, each row representing the term statistics of both a German and an English Wikipedia article on the same topic. The German monolingual block contains the German term statistics of the remaining 1000 Wikipedia articles. Likewise, the English monolingual block contains 1000 articles, each with their counterpart in the German monolingual block.

The evaluation setup is as follows: For every document $d^{(de,i)}$ in the German monolingual block we rank the document in the English monolingual block according to the retrieval method applied and get a rank $r_i$ for the one corresponding to the German query document. A run is evaluated by its mean reciprocal rank (mrr), given by:

$$mrr = \frac{1}{N} \sum_{i=1}^{N} \frac{1}{r_i}$$

For the pLSA model, the whole matrix of three blocks is embedded, that is, the $P(Z|d_i)$ of both the comparable and the monolingual articles were computed and used in the re-estimation iterations. This setting might be unrealistic in a real-word retrieval task, however as pLSA does theoretically not define inference for unseen documents it seems to us to be the most sound way. Note that in this scheme also term co-occurrences in the monolingual blocks influence the model estimation. For every training run, 100 EM iterations were performed.

Ranking was done by the cosine similarity of the parameter vectors of the multinomial distributions $P(Z|d_i)$, a normalized dot product lying between 0 and 1 for vectors with all positive components:

$$sim(d^{(1)}, d^{(2)}) = \frac{\sum_{k=1}^{K} P(z_k|d^{(1)}) P(z_k|d^{(2)})}{\sqrt{\sum_{k=1}^{K} P(z_k|d^{(1)})^2} \sqrt{\sum_{k=1}^{K} P(z_k|d^{(2)})^2}}$$

This ranking scheme was applied and found to work best by Hofmann [2001] for a monolingual retrieval task and standard pLSA.

Some experiments with pLSA involve a weighting of the data matrix. Only weighting schemes were applied that changed the length of the documents (to which $p(d_i)$ is directly proportional) and therefore their influence. Such a scaling is theoretically admissible as it does not touch the meaning of the other concepts and can give hints which data provide the most useful cross-language bridge.

---

[6] http://snowball.tartarus.org/

For the ESA model we applied the parametrization described in 3.3 and used the comparable block as a training corpus. Note that the cut-off parameter is greater than the training corpus size and has therefore no influence. Again ranking is done by the cosine similarity.

For the LDA model we estimated the parameters $P(W|Z)$ with the comparable block. The highly efficient and parallel LDA implementation *plda* [Wang et al., 2009] was used. We used the Dirichlet parameter values suggested by Griffiths and Steyvers [2004], $\alpha = \frac{50}{\text{(number of topics)}}$ and $\beta = 0.1$. The model was trained by doing 100 sampling iterations in order for the distributions to converge, and then averaging the sampling outcomes of further 50 iterations for the estimates of $P(W|Z)$. For inference on unseen documents, 10 iterations were used for convergence and further 5 iterations for averaging the estimate. German and English terms were made distinguishable by adding the prefixes *de_* and *en_* respectively. Inference was done with this model on the documents in the monolingual blocks. Ranking was made, in the same way as for pLSA and ESA, by taking the cosine of vectors of the sampled topic counts per document.

### 4.1.2 Results

In the first runs we tested the multilingual pLSA model. The glue documents are the only cross-language bridge and therefore of special importance. Instead of varying the number of glue documents (which is 1000), we gave them higher weight. Both single runs as well as combined runs with concatenated parameter vectors of different dimensionalities were evaluated. Combination was only done for runs with the same weighting.

Model combination has a major impact on performance, which is explainable by overfitting issues and convergence to local maxima. The first applied weighting scheme multiplied the term counts in the comparable block by a constant weight (we tried values of 2,4,8 and 16). As Figure 6 and Table 2 show, this improved retrieval performance drastically compared with the original model. The higher the weights are chosen, the more the $P(w|z)$ are influenced by the comparable documents, the monolingual doc-uments are then rather "pulled" in the latent spaces instead of "pulling" themselves. However, this is only true for the $P(w|z)$ that occur also in the comparable documents. The $P(w|z)$ for words that occur only in monolingual documents provide some transitive clues that would not be present if the model was trained on comparable data only.

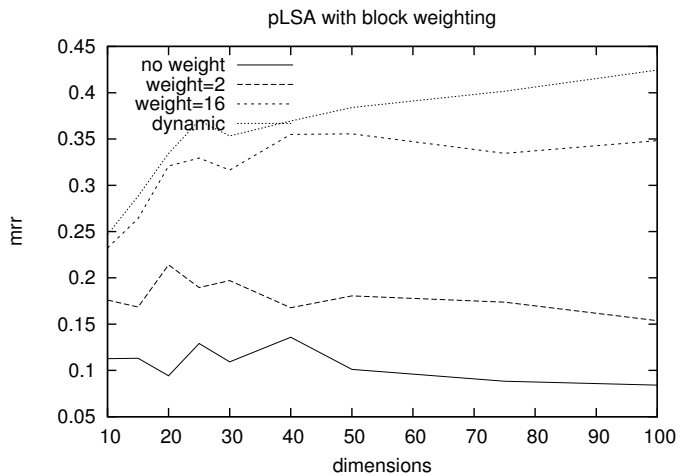

Figure 6: Weighting of the multilingual block of the data matrix. Increasing the weight of the glue documents boosts performance. Even better is a weight that favors equally long documents in both languages.

Table 2: Combination of pLSA runs with number of topics $= 10, 25, 50, 75, 100$.

| weighting | mrr |
|---:|:---|
| w/o | 0.2767 |
| weight $= 2$ | 0.4427 |
| weight $= 4$ | 0.5740 |
| weight $= 8$ | 0.6120 |
| weight $= 16$ | 0.6157 |
| dynamic | 0.6551 |

The other weighting scheme is based on the reasoning that, instead of uniformly weighting multingual documents higher, it could be beneficial to enforce documents that are more multilingual than others. One criterion to regard documents as more multilingual is the degree to which they are similar in length in both language parts. Therefore we formalize the weight as the geometric mean, dependent on the length of the two language sides $l_{de}$ and $l_{en}$ with a small floor count $c = 0.001$. We call such a document weighting "dynamic". The dynamic weight yielded the best results for pLSA, it is formalized as:

$$w_{dyn} = \sqrt{(1-c)\frac{l_{de}}{max(l_{de}, l_{en})} + c} \cdot \sqrt{(1-c)\frac{l_{en}}{max(l_{de}, l_{en})} + c}$$

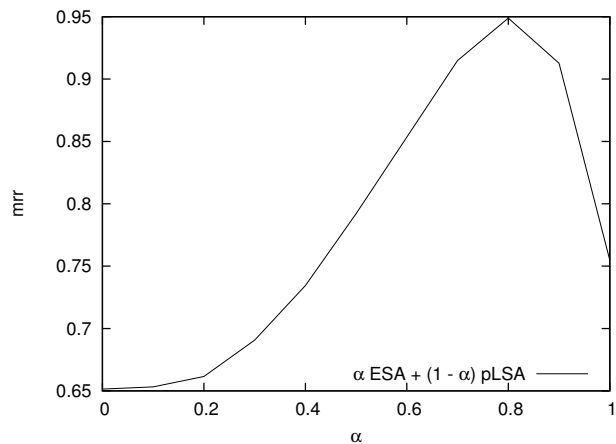

Figure 7: Performance of the linear combination of the ESA and pLSA (dynamic weight) vectors.

With the ESA model we got a score of $mrr = 0.7548$, which is clearly better than the pLSA model. In order to test whether the ESA model conveys distinct information to the pLSA model, we interpolated both models and multiplied the ESA vectors by a parameter $\alpha$ and the dynamically weighted pLSA vectors by $(1-\alpha)$ and used the concatenation of both. We got an improvement by 25.7% with $mrr = 0.9489$ for $\alpha = 0.8$. See Figure 7.

For LDA we performed three types of experiments: First, we sampled a model for the comparable documents with standard LDA without any length normalization, in the manner of [Cimiano et al., 2009]. In fact, it turned out that this produced the worst results of all runs conducted. To estimate whether the different length ratios were responsible for the negative effect, we constructed a new training set containing the same information, but with same length ratios. For this end we multiplied the counts of the shorter language side to match the counts of the longer one. As the sampling procedure needs integers, we rounded the new counts. Figure 8 and Table 3 show that this has a drastic impact on performance. This is in accordance with our theoretical assumptions. In another experiment on the stretched data, we multiplied (after inference was done) each sampled topic outcome with a weight according to the inverse document frequency of the word it was sampled for. The reasoning behind that was to increase the influence of discriminative terms. We found that this brought a slight improvement but it does not preserve the theoretical properties of the topic statistics.

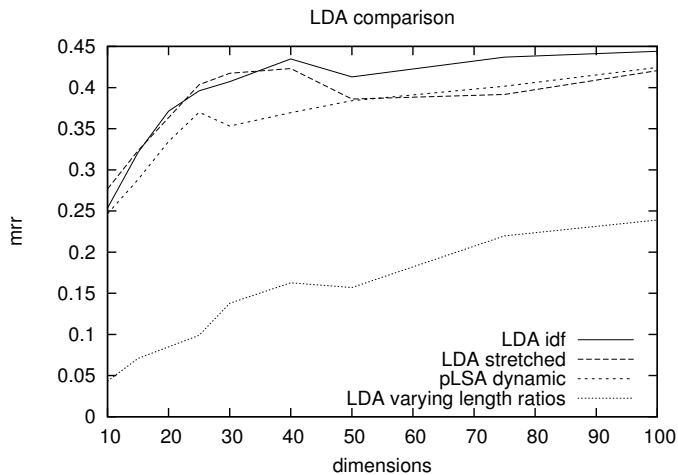

Figure 8: Latent Dirichlet Allocation on the Wiki subset. The not length-normalized version performs worst.

Table 3: Combination of LDA runs with number of topics $= 10, 25, 50, 75, 100$.

| model | mrr |
|---|---|
| LDA varying length ratios | 0.1960 |
| LDA stretched | 0.6307 |
| LDA idf | 0.6539 |

In order to estimate how much different information the two latent variable models convey, we interpolated the dynamically weighted pLSA ($\alpha$) model with the idf weighted stretched LDA model ($1 - \alpha$). We got $mrr = 0.7063$ for $\alpha = 0.4$, which is an improvement of 8.0% with respect to the LDA run alone.

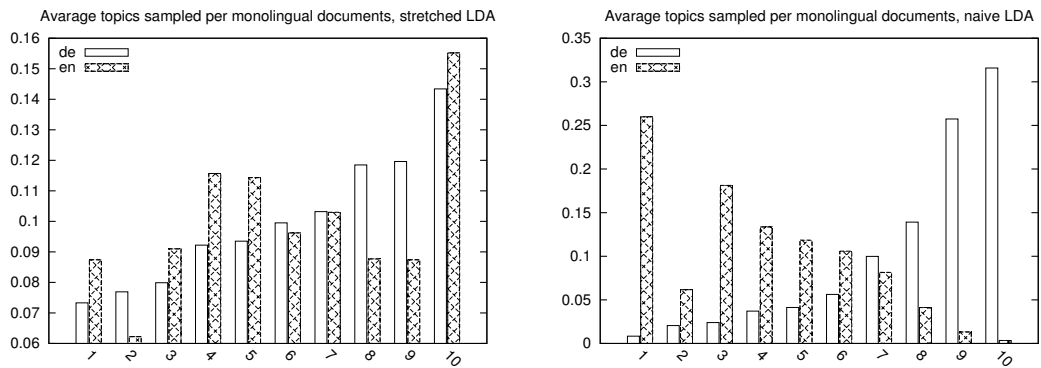

Figure 9: Relative frequency of inferred topics for 10 dimensional LDA models, sorted by relative frequency for the German monolingual (held-out) documents. The unstretched model (right) shows clear negative correlation for German and English sampling statistics.

Figure 9 shows topic distributions for the German and the English monolingual blocks for which inference was done with LDA models trained on either the stretched or the bare comparable block. It can never be expected that the inferred topics are the same for two text sets; in our case the monolingual documents might focus on different thematical aspects. However, for the unstretched model it is almost consistently the case that topics predominant in one language do rarely occur in the other language. We take this as evidence that such a model primarily estimates the language of a document.

We also tried runs that applied a tf.idf weighting to the document matrix in the pLSA setting. This weighting had a slightly negative effect on performance. Moreover, it does not integrate well in the theoretical models, therefore we will not discuss these results further.

### 4.1.3 Conclusion

The initial experiments have been made on rather a small training and target collection. However, the number of query documents is high and therefore some interesting observations can be made. Obviously it is important to save model capacities for cross-language bridging. We can assume that performance improves

- when the model does not try to structure the monolingual spaces. The indicator

for this is the downweighting experiment for imbalanced documents in the pLSA model.

- when the model does not try to guess the language. The upscaling experiment of the shorter language side points to this.

Furthermore, performance increases drastically when different runs of latent variable models are combined and when latent variable models are interpolated with ESA.

## 4.2 Mate Retrieval on Multext JOC

---

**Q: Welche Änderungen schlägt die Kommission vor, um mit einem ermäßigten Steuersatz Anreize für die verstärkte Entfernung von Gestrüpp aus den Wäldern zu bieten? A: Die Frage, wie hoch die Mehrwertsteuer für Dienstleistungen in der Land- und Forstwirtschaft ab 1993 sein soll, wird zur Zeit im Ministerrat diskutiert.**

---

**Q: What tax changes does the Commission intend to introduce to encourage clearance of undergrowth? A: The question of the VAT rate to be applied to agricultural and forestry services from 1993 onwards is currently under discussion in the Council.**

---

Figure 10: Shortened question and answer pair of the Multext JOC corpus.

The Multext JOC corpus[7] consists of 3500 questions to the European Parliament and of answers to these questions. As in Sorg and Cimiano [2008] we use the concatenation of a question together with its answer as a query in one language to search the collection of translations in another language for its counterpart. Our experiments were done with English as the query language and German as the target language. Only preprocessing steps that are clear and easy to reproduce were performed. Exactly those questions were retained that to which an answer was assigned and had the same id in English and German. This resulted in a set of 3212 texts in each language, 157 more than were used in Sorg and Cimiano [2008][8]. Sequences of characters in Unicode letter blocks were considered words. Words with length = 1 or length > 64 and words contained in the Snowball stopword list were ignored. All other words were stemmed with the publicly available Snowball stemmer[9]. In contrast to Sorg and Cimiano [2008], no compound splitting was done.

For the training collection all pairs of Wikipedia articles[10] were used that have bidirectional direct cross-language references. All markup was stripped off by using the same filter as in a publicly available ESA implementation[11]. Wikipedia articles of less than 100 words in either language were ignored and words with a Wikipedia document frequency of 1 were filtered out. The final training corpus consists of 320000 bilingual articles.

---

[7] http://www.lpl.univ-aix.fr/projects/multext
[8] the exact document selection criterion of their experiments is unknown to us
[9] http://snowball.tartarus.org/
[10] we used the German snapshot of 2009-07-10 and the English snapshot of 2009-07-13

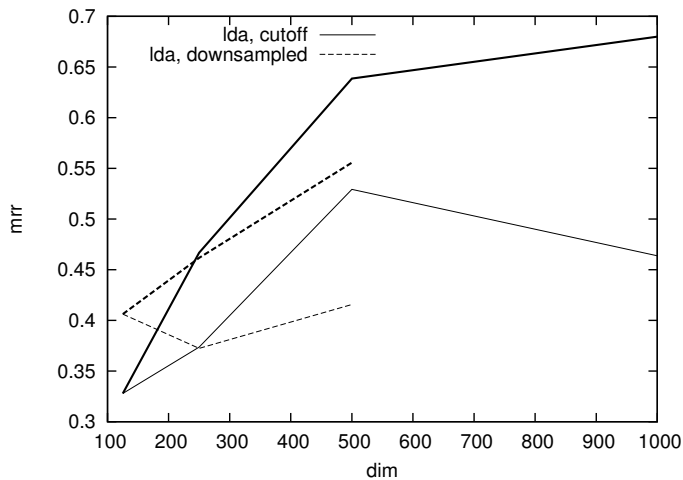

Figure 11: Performance of LDA models estimated with dimensions (=numbers of topics) equal to 125, 250, 500 and 1000. Thick lines indicate combinations of all models up to the dimension on the $x-$axis. Drops in the perfomance curve on single points may be due to local sampling optima.

Performance of retrieval was measured in mean reciprocal rank (mrr). The ESA retrieval experiment was performed using the same parametrization as discribed before and the result of Sorg and Cimiano [2008] was reproduced to a difference of 1% (in our experiments we obtained a score of $mrr = 0.77$ compared with $mrr = 0.78$).

As for the LDA experiments, we were interested in the effect of length normalization of the training documents. We tried two methods: First, every document was cut off at a length of 100 words. Second, for the longer language side of an article only a random sample of size equal to the smaller language side was retained. A resizing with a scalar is not possible because the sampling process requires integer counts. We marked each word with a prefix indicating its language and retained a vocabulary size of roughly 770 thousand and 2.1 million for the cut-off method and for the downsampling method respectively. Both training collections were embedded with 125, 250 and 500 dimen-sions, and additionally with 1000 dimensions for the cut-off corpus (the vocabulary size was the limiting factor with respect to our computing facilities). The Google plda package Wang et al. [2009] was used with the suggested parameters ($\alpha = \frac{50}{\#\text{topics}}$ and $\beta = 0.01$). With the trained model, topics were inferred for the monolingual Multext documents. In order to get a stable estimate, the statistics of 50 sampling iterations were averaged. Similarity in the LDA setup was measured by taking the cosine similarity between the sampling statistics.

---
[11] http://code.google.com/p/research-esa/

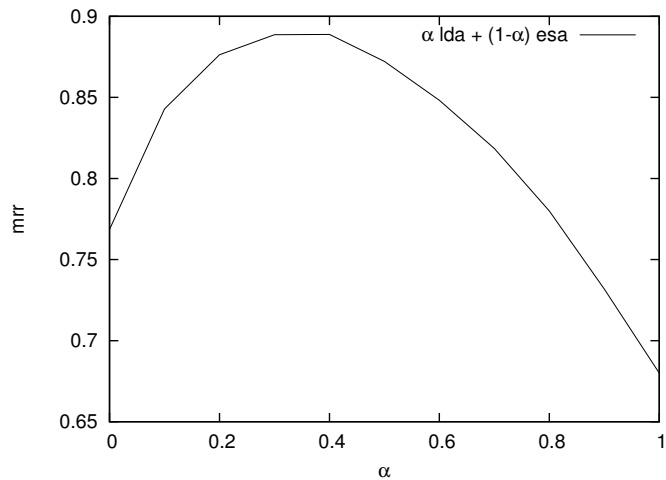

Figure 12: Combining LDA and ESA on the Multext corpus. The improvement over ESA alone is significant with $p \ll 0.005$ for $.1 \leq \alpha \leq .7$.

Table 4: Performance of LDA on Multext

| LDA method | number of topics | mrr |
|---|---|---|
| Sorg & Cimiano | 500 | .16 |
| length downsampling | 500 | .42 |
| length cut-off | 500 | .53** |
| length downsampling | 125 + 250 + 500 | .55** |
| length cut-off | 125 + 250 + 500 + 1000 | .68** |

A drastic improvement over non-normalized LDA can be observed: while Sorg and Cimiano [2008] report a score of $mrr = 0.16$ for their 500-dimensional LDA model, we get $mrr = 0.53$ with the cut-off corpus. We suppose that the reason for this difference is that a non-multilingual LDA model applied to a comparable corpus estimates the predominant language of a document rather than its semantic content. Another improvement can be observed by combining the results of different models, a technique that is usually applied for pLSA Hofmann [2001]. In this case, the cosine scores of runs with different dimensional models were simply averaged (this corresponds to concatenating the $L_2$-norm normalized sampling statistics vectors). This yielded a score of $mrr = 0.68$ for the cut-off model, showing performance in the same order of magnitude as ESA. Figure 11 and Table 4 give a survey of the results obtained with LDA. Scores significantly better than in the respective line above having $p \ll 0.005$ in the paired t-test are marked with $**$. (Of course we could not test against scores reported elsewhere, for lack of the original numerical data.)

How different are the ESA and the LDA models, how much can they contribute to each other? In order to answer this question, we combined the cosine scores of both models by different interpolation factors $0 \leq \alpha \leq 1$. A stable improvement in performance with maximum $mrr = 0.89$ was achieved for giving the cut-off LDA model a weight of 0.4 and the ESA model a weight of 0.6. See Figure 12.

> **Average Precision** measures for every document $i$ relevant to a query $q$ ranked at $r_{i,q}$ the precision up to its rank. The per-document score
> $$AP_{i,q} = \frac{\text{relevant documents } j \text{ with } 1 \leq r_{j,q} \leq r_{i,q}}{r_{i,q}}$$
> is combined to a per-query score, by averaging over all scores of the set of relevant documents $Rel_q$:
> $$AP_q = \frac{1}{|Rel_q|} \sum_{i \in Rel_q} AP_{i,q}$$
> **Mean Average Precision** combines these scores for the set $Q$ of all queries by their arithmetic mean:
> $$MAP = \frac{1}{|Q|} \sum_{q \in Q} AP_q$$
> **Geometric Mean Average Precision** uses the geometric mean instead:
> $$GMAP = \sqrt[|Q|]{\prod_{q \in Q} AP_q}$$

## 4.3 Query-Based Retrieval with CLEF2000

### 4.3.1 Experiment Setup and Results

Mate retrieval experiments can be criticized as being an unrealistic retrieval scenario. Therefore, a second evaluation was done on the CLEF[12] German-English ad-hoc track of the year 2000. The target corpus consists of about 110000 English newspaper articles together with 33 German queries for which relevant articles could be pooled. For our experiments the title and description fields of the queries were used and the narrative was ignored.

A common strategy for cross-language retrieval is first to translate the query and then to perform monolingual retrieval. While the translation process would have taken prohibitively long for the Multext corpus, we performed query translation on the CLEF2000 queries with a standard Moses translation model trained on Europarl. Retrieval with the translated queries was done by comparing the cosine of the tf.idf-weighted word-vectors; we used formula (11) on page 29 to compute this weighting.

We evaluated both the machine translation model and the concept models trained on

---

[12]http://www.clef-campaign.org/

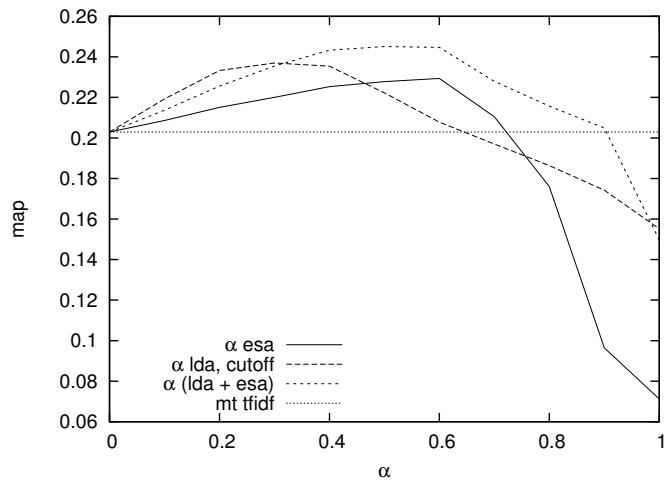

Figure 13: Interpolation of concept models with machine translation on CLEF2000. The improvement given by the LDA-ESA is significant with $p < 0.005$ for $.2 \leq \alpha \leq .6$.

Wikipedia. In addition to the most commonly used mean average precision score (*map*) we also evaluated by geometric mean average precision (*gmap*, see page 46), which rewards stable results for hard queries. The ESA and the cut-off LDA models with $dim = 500$ perform equally well for *map*, while the combination of LDA dimensions gets a considerably better score. This is in contrast to the findings in the mate-retrieval setup. The reason for that, we suspect, may be that the parameters of ESA have been found in order to optimize such a setting. For *gmap*, LDA consistently outperforms ESA. For the combined LDA-ESA model no clear improvement could be observed using the combination with $\alpha = .4$ from the previous experiment. The machine translation model ($map = .203$) performed better than the concept models. When the different concept models were combined with the machine translation model by interpolating their cosine scores, all three models achieved improvements. The biggest and most stable improvement was achieved by the LDA-ESA concept model yielding a score up to $map = .246$ with equal weight for the concept model and the machine translation model. Figure 13 and Table 5 show an overview of the results. Scores significantly better than in the respective line above, having $p < 0.05$ and $p < 0.005$ in the paired t-test, are marked with $*$ and $**$ respectively.

Table 5: Query-based retrieval on CLEF2000

| method | parameters | map | gmap |
|:---:|:---:|:---:|:---:|
| ESA | | .071 | .003 |
| LDA, cut-off | $d = 500$ | .071 | .010 |
| LDA, cut-off | $d = all$ | .155* | .043* |
| LDA+ESA | $\alpha = .4$ | .150 | .051 |
| MT tf.idf | | .203 | .061 |
| concept+MT | $\alpha = .5$ | .246** | .125** |

### 4.3.2 Error Analysis

A querywise error analysis is difficult because the inner workings of quantitative methods are often opaque to human analysis. However, the machine translation output is the most contributing source and it is accessible to examination. We sorted the machine translation output by how much it profited by the concept models in the best performing setting. On page 50 we report the score that is obtained by machine translation and the increase when combined with the concept models. We analyzed how often a word was obviously unknown by the machine translation system trained on Europarl and therefore wrongly just copied over. It would be possible to recognize this type of error automatically. In addition, for every translated query we counted how many words in it had no semantic meaning related to the purpose of the query and were therefore useless (these words are hence called *junk words*). Junk words are, for example, function words not filtered by the stopword list, machine translation errors of several kinds and artefacts from the query formulation, mainly from the description part (e.g. "relevant documents contain information about..."). The junk word error type would be more difficult to detect.

Although the analyzed data basis is small, we conjecture that the concept model makes such queries more robust which induce one of the two errors, while it might be less useful where a good translation is present and the terms are weighted well: In the cases where the concept models contributed there were, on average, 0.53 unknown words for machine translation and 0.24% junk words, in contrast to 0.14 unknown words and 0.04% junk words in the cases where the concept model decreased the score. For future experiments it might be interesting to test whether a trade-off weighting between translation and concept model conditioned on a reliability score of the machine translation improves performance.

## 4.4 Discussion

We have shown that a highly significantly contributing cross-language retrieval component can be obtained with LDA, since standard and multilingual LDA behave alike for length normalized data. The two very simple techniques we found most effective are document cut-off and model combination. Thus, we get an improvement of 325% $mrr$ compared with a non-competitive score previously reported for LDA on the Multext corpus. The combination of LDA with ESA increases performance in mate-retrieval experiments by 15.6% $mrr$ compared with ESA alone. In cross-language query-based retrieval on CLEF 2000, a monolingual scheme based on machine translation output can be improved by 21.1% $map$ when combined with concept models. While ESA alone performs better on mate-retrieval, we find LDA superior in query-based retrieval, maybe because commonly used ESA-parameters have been tuned for direct translation finding.

|    |                                              | scores   |              | errors    |         |
|----|----------------------------------------------|----------|--------------|-----------|---------|
| id | query title                                  | mt *map* | $\Delta map$ | unknowns  | junk(%) |
| 18 | Unfälle von Brandbekämpfern                  | 0.000    | $map = 0.033$ | +        | 0.71    |
| 9  | Methanlagerstätten                           | 0.000    | $map = 0.017$ | ++       | 0.50    |
| 7  | Doping und Fußball                           | 0.004    | +361.36%     |           | 0.33    |
| 4  | Überschwemmungen in Europa                   | 0.007    | +209.09%     |           | 0.17    |
| 26 | Nutzung von Windenergie                      | 0.071    | +167.40%     |           | 0.13    |
| 11 | Neue Verfassung für Südafrika                | 0.093    | +111.95%     |           | 0.09    |
| 12 | Sonnentempel                                 | 0.009    | +109.57%     | +         | 0.50    |
| 17 | Buschbrände bei Sydney                       | 0.138    | +96.09%      | ++        | 0.29    |
| 5  | Mitgliedschaft in der Europäischen Union     | 0.060    | +89.38%      |           | 0.08    |
| 15 | Wettbewerbsfähigkeit der europäischen...     | 0.177    | +88.51%      |           | 0.10    |
| 38 | Rückführung von Kriegstoten                  | 0.045    | +86.84%      | +++       | 0.33    |
| 21 | Europäischer Wirtschaftsraum                 | 0.165    | +73.38%      |           | 0.22    |
| 14 | USA-Tourismus                                | 0.015    | +67.72%      | +         | 0.17    |
| 28 | Lehrmethoden für nicht-englischsprachige...  | 0.072    | +61.43%      | +         | 0.29    |
| 33 | Krebsgenetik                                 | 0.382    | +52.46%      | +         | 0.38    |
| 16 | Die Französische Akademie                    | 0.084    | +46.84%      |           | 0.33    |
| 13 | Konferenz über Geburtenkontrolle             | 0.137    | +46.68%      |           | 0.17    |
| 40 | Privatisierung der Deutschen Bundesbahn      | 0.015    | +46.10%      |           | 0.25    |
| 31 | Verbraucherschutz in der EU                  | 0.222    | +46.03%      |           | 0       |
| 32 | Weibliche Priester                           | 0.234    | +34.99%      |           | 0.22    |
| 20 | Einheitliche europäische Währung             | 0.218    | +24.13%      |           | 0       |
| 36 | Produktion von Olivenöl im Mittelmeerraum    | 0.269    | +23.29%      |           | 0.11    |
| 22 | Flugzeugunfälle auf Start- und Landebahnen   | 0.094    | +19.76%      | +         | 0.22    |
| 37 | Untergang der Fähre Estonia                  | 0.926    | +4.13%       |           | 0.17    |
| 19 | Golfkriegssyndrom                            | 0.704    | +1.44%       |           | 0.25    |
| 1  | Architektur in Berlin                        | 0.651    | +1.27%       |           | 0.17    |
| 30 | Einsturz einer Supermarktdecke in Nizza      | 1.000    | −.01%        | +         | 0.13    |
| 24 | Welthandelsorganisation                      | 0.516    | −6.94%       |           | 0       |
| 10 | Krieg und Radio                              | 0.015    | −11.47%      |           | 0       |
| 29 | Erster Nobelpreis für Wirtschaft             | 0.072    | −11.99%      |           | 0       |
| 39 | Investitionen in Osteuropa oder Rußland      | 0.041    | −16.99%      |           | 0.15    |
| 34 | Alkoholkonsum in Europa                      | 0.060    | −24.18%      |           | 0       |
| 3  | Drogen in Holland                            | 0.185    | −55.66%      |           | 0       |

Table 6: Improvement of *map* through the ESA+LDA concept component compared with error rates. The 4th column indicates the change in comparison to machine translation alone (3rd column). The 5th column contains one + per unknown word, the last column the percentage of junkwords per query.

# 5 Towards Retrieval Based on Only Wikipedia

Concept models trained on article level can contribute to cross-language retrieval. However, the main contribution comes from the statistical machine translation module trained on Europarl. Coarse thematical relationships alone can arguably not capture the specific meaning contained in a query, a word-to-word translation is necessary. But lexica and parallel corpora are often only accessible for some European language pairs. And even if they exist their vocabulary is inherently limited in contrast to an ever growing ressource like Wikipedia. The question therefore arises how parallel text or word-to-word mappings that are less dispersing than those obtained from document-level co-occurrences can be replaced by information extracted from freely available knowledge sources such as Wikipedia.

Several approaches in this direction have been undertaken. Often, a Wikipedia title in the source language is associated with a word and the corresponding title in the target language is used as a translation [Wentland et al., 2008, Nguyen et al., 2009]. However, Sjobergh et al. [2008] note that the vocabulary distribution of titles has a skew to certain words: 33% are normal nouns, 66% proper nouns and only 0.03% verbs. Because the titles of Wikipedia articles are specifically tailored to be unique identifiers rather than representative text samples, translations of large classes of words might introduce artefacts. Also, titles are usually in a normal form which could be problematic in highly inflecting languages or when a variation is used. Another, more general approach is to construct a parallel corpus out of a comparable corpus by finding parallel sentences [Maeda et al., 2008, Adafre and de Rijke, 2006] which can then be used to enhance the training set of a machine translation system. However, such an extraction is complicated and still requires a parallel corpus to perform well.

An optimal translation model based on Wikipedia should therefore fulfill the following demands:

- It should be based on simple ideas and easily implemented.

- It should be based on probabilistic principles to be usable in a language modeling setup.

- The model should be text-to-text rather than title-to-title.

- The model should only translate text it is trained for.

The method we use is based on the anchor text of links. In Wikipedia, text that is expected to be of particular interest to a reader is linked to another Wikipedia article if this adds to the understanding. The key policy as stated in the Wikipedia guidelines is:

*Ask yourself, "How likely is it that the reader will also want to read that other article?" These links should be included where it is most likely that readers might want to use them; [...] Always link to the article on the most specific topic appropriate to the context from which you link: it will generally contain more focused information, as well as links to more general topics.*[13]

Any text can be linked to any page. For example, the German texts "in Wäldern gelegte Brände", "Buschfeuer", "Buschbrand" and "Waldbrand" may be valid contexts to be linked to the article titled "Waldbrand" (with the English counterpart "Wildfire"). The following three items of information are necessary to build a (potentially multilingual) probabilistic translation model based on linked text:

- How likely is a text to be linked?

- What are probable links, given a linked text?

- What is a probable text, given there is a link to a certain article?

## 5.1 Model

If, for the sake of simplicity, a unigram model is used this amounts to the probablities $P(l|w)$, $P(a|w,l)$ and $P(w|a,l)$. In a bilingual setting $l$ is a variable indicating whether the source word is linked, $a$ is the bilingual article the source word is linked to, and $w_E$ and $w_F$ are words in the source and the target languages respectively. It is:

$$\begin{aligned}P(w_F|w_E) &= P(w_F|l=\text{true}, w_E)P(l=\text{true}|w_E) \\ &\quad + P(w_F|l=\text{false}, w_E)P(l=\text{false}|w_E)\end{aligned}$$

Assuming that the translation of linked words does only depend on the articles they are linked to, one gets:

---

[13]http://en.wikipedia.org/wiki/Wikipedia:Linking retrieved on 13 October 2009

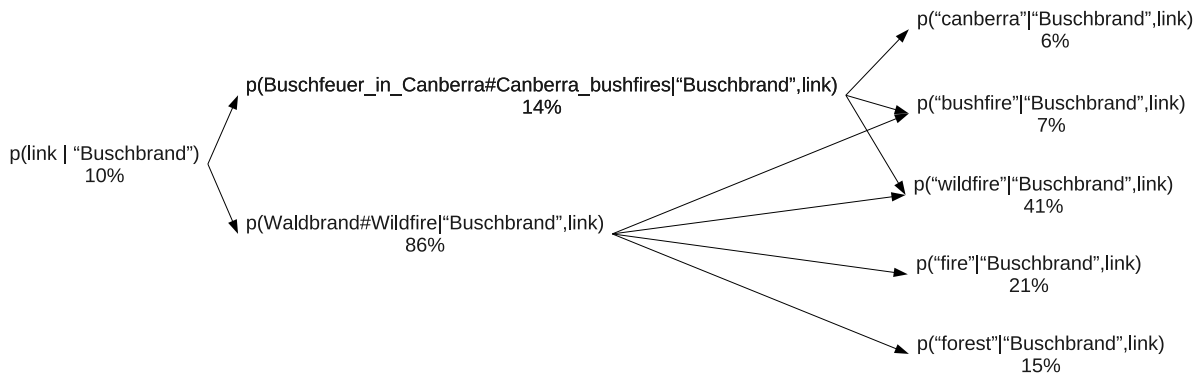

Figure 14: Example values of the components of the link translation model.

$$P(w_F|l = \text{true}, w_E) = \sum_a P(w_F|a, l = \text{true})P(a|l = \text{true}, w_E)$$

If Wikipedia links are the only cross-language bridge, for the unlinked case a general language model $LM_F$ could be used to estimate the probability, also other translation sources could be plugged in here.

$$P(w_F|l = \text{false}, w_E) = P(w_F|LM_F)$$

In [Mihalcea and Csomai, 2007] several methods were tried to predict whether a word was linked or not. The most effective was based on the relative frequency of linked occurrences. We take this approach also for the other distributions and get

$$P(l = \text{true}|w_E) = \frac{n(l = \text{true}, w_E)}{n(w_E)}$$
$$P(a|l = \text{true}, w_E) = \frac{n(a, w_E)}{n(l = \text{true}, w_E)}$$
$$P(w_F|a, l = \text{true}) = \frac{n(w_F, a)}{n(a)}$$

where $n(\cdot)$ denotes co-occurrence counts and we omitted redundant restrictions. Figure 14 shows some values for an example translation. Note that the translation distribution is not as clean as one could whish (the city name "Canberra" is linked as strongly associated with a specific event of a wildfire). One might hypothesise that problems of that kind are connected with the unigram assumption: If the anchor texts

---

**C001** architektur 2 gesucht 1 berlin 2 dokument 1

**C003** niederland 1 holland 1 drogenpolit 1 drog 1

**C004** lief 1 zahl 1 verursacht 1 uberschwemm 2 gesucht 1 wirtschaft 1 europa 2 dokument 1 zerstor 1 kost 1 landwirtschaft 1

---

**C001** architectur 19 berlin 49 univers 1

**C003** drug 23 south 1 dutch 9 pharmaceut 2 polici 12 netherland 17 holland 15

**C004** european 9 parliament 1 econom 1 economi 1 eu 1 agricultur 15 number 1 union 3 europ 16 flood 3 cost 2 destroy 9

---

Figure 15: Translations from German to English stemmed bag-of-word queries delivered by the translation model, for the first three queries. Counts in the translated Texts denote $100 \cdot P(w_{\text{transl}}|l = \text{true}, Q_{\text{orig}})$.

were considered as a whole, the specifity of the articles would correspond better to the specifity of texts.

Many ways of how to exploit such a model are conceivable. For example, it can be used for query translation. First we note that the probability $P(l = \text{true}|w_E)$ can function as a term weighting, assuming that the importance of terms is correlated with their probability of being linked. We use this assumption and employ $P(w_F, l = \text{true}|w_E)$ for a translation model. It is important to keep in mind that $l$ indicates the linking of the *source* word. Additionally, translation and linking should be independent of the document, given a source word (but not independent of each other). The translation model of a document $D_E$ is the probability distribution that can be paraphrased as:

*The probability that a word, chosen from D, is linked and has $w_F$ as translation.*

Using the parts defined before and simple reordering gives:

$$\begin{aligned} P(w_F, l = \text{true}|D_E) &= \sum_{w_E} P(w_F, l = \text{true}|w_E) P(w_E|D_E) \\ &= \sum_a P(w_F|a, l = \text{true}) \sum_{w_E} P(a|l = \text{true}, w_E) P(l = \text{true}|w_E) P(w_E|D_E) \end{aligned}$$

Note that the inner sum defines $P(a, l = \text{true}|D_E)$, which similarly to ESA associates a weighting over Wikipedia articles to documents but in addition has a probabilistic

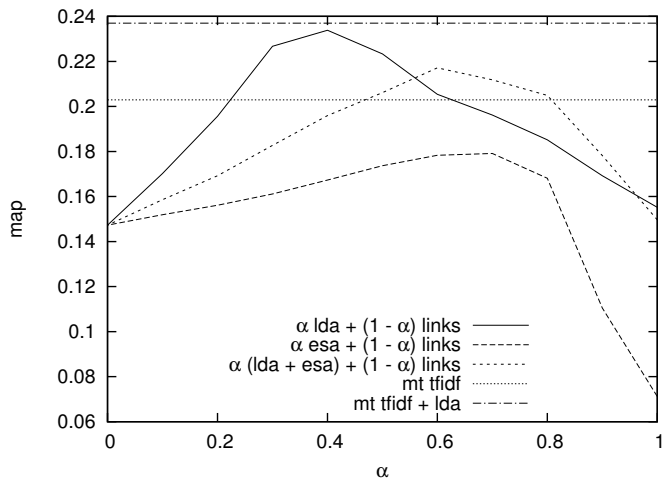

Figure 16: Retrieval results of the tf.idf weighted link translations, combined with different concept models.

semantics. By normalization, $P(w_F|l = \text{true}, D_E)$ can be obtained easily.

## 5.2 Vector Space Experiments

Experiments are done on the CLEF2000 dataset. The queries are translated from German into English by the following scheme: In the translation of the original query $Q_{\text{orig}}$, words $w_{\text{transl}}$ with $P(w_{\text{transl}}|l = \text{true}, Q_{\text{orig}}) \geq 1$ are retained. Counts are assigned to them proportionally to that probability. Retrieval is done with the translated query by tf.idf weighting and taking the cosine in the same way as it was done for queries translated with Europarl. Additionally, we interpolated the cosine scores of this run with the scores obtained by the other models trained on Wikipedia, which comprise the LDA model, the ESA model and the combination of LDA and ESA that worked best in the Multext experiments of section 4.2. In combination with the link translation model, LDA showed the best and most stable performance of up to $map = 0.234$. This is better than retrieval with the Moses translated queries ($map = 0.203$). However, it is slightly worse than the combination of machine translation output with LDA (0.237).

ESA was clearly the least contributing component and could not improve the combination of the link model with LDA. One explanation may be that ESA captures similar information as the link model but does so less successfully. LDA and the link model, however, seem to be complementary, leading to a huge improvement when they are combined. As an alternative way to make use of the link model, similarly to the ESA model, the cosine similarity was taken for vectors with association strengths to Wikipedia articles. In the link model, we used the distribution vector $P(a|l = \text{true}, Q_{\text{orig}})$ and $P(a|l = \text{true}, D)$ to establish a similarity between query $Q$ and document $D$. Using the article distribution vectors was a less successful way of employing the link model, but still better than the ESA model. Figure 16 and Table 7 give a survey of the results obtained with the link translation model.

Table 7: Query translation based on links (CLEF2000). Moses scores are given for comparison.

| method | parameters | map | gmap |
|---|---|---|---|
| link probability vector | | .093 | .013 |
| link translation | | .147 | .043 |
| link translation + ESA | $\alpha = .7$ | .179 | .031 |
| Moses | | .203 | .061 |
| link translation + LDA | $\alpha = .4$ | .234 | .110 |
| Moses + LDA | $\alpha = .3$ | .237 | .124 |

## 5.3 Language Modelling Experiments

In a query likelihood model [Ponte and Croft, 1998] a document provides a probabilistic model for a query. The ranking is usually done by:

$$\log P(Q|D) = \sum_{w \in Q} n(w, Q) \log P(w|D)$$

In principle, all components are provided by the link model to perform retrieval in such a fashion. The probability that article $a$ is the target when a linked word is picked from document $D$ (making the same independence assumptions as before) is

$$P(a|D, l) = \frac{\sum_{w_E} P(a|l = \text{true}, w_E) P(l = \text{true}|w_E) P(w_E|D)}{\sum_{a', w_E} P(a'|l = \text{true}, w_E) P(l = \text{true}|w_E) P(w_E|D)}$$

and $P(w_F|a, l)$ is the probability that, given a source word linked to article $a$, it is translated to word $w_F$. Together the elements of the link model provide us with the

distribution:
$$P_{link}(w_F|D,l) = \sum_a P(w_F|a,l)P(a|D,l)$$

It is practical to think of this distribution as combined in that way, because it separates the per-document estimates from the vocabulary estimates. Also, it is important to remember that the distribution is only conditioned on the words expected to be linked in the source document (indicated by the conditioning on $l$). If one wants to use the model as it is for a query likelyhood model, three problems immediately turn up:

1. **The zero-probability problem:** Because the components of the model are estimated from relative frequencies, to many events zero probability is assigned. If one word in a query has zero probability, the whole query has also.

2. **The complexity problem:** If the probability distributions are smoothed so that they are never zero, summation might for every word go over all Wikipedia articles, which would be inhibitively expensive to compute.

3. **The training basis problem:** The model only considers words likely to be linked. Especially high frequency or function words could have skewed distributions.

### 5.3.1 Model Combination on Query Level

The zero-probability problem is especially severe when

$$P_{link}(w_F|D,l) = \sum_a P(w_F|a,l)P(a|D,l) = 0$$

because this makes the log-probability undefined. This happens when $P(w_F|a,l) = 0$ for the articles $a$ that have non-zero probability for document $D$. However, smoothing a probability $P(w_F|a,l) > 0$ for all word and document combinations would result in a full matrix with number of entries equal to vocabulary size multiplied by Wikipedia size. Moreover, smooth distributions $P(w_F|a,l)$ and $P(a|D,l)$ would require full summation over all articles, while in the sparse case only the matching non-zeroes have to be considered. We employ a simpler, although slightly less principled strategy. We floor the probability $P_{link}(w_F|D,l)$ by the minimal value that could have been obtained by smoothing $P(w_F|a,l)$ with Laplace (add-one) smoothing.

$$P_{link}(w_F|D,l) = \max\left(\sum_a P(w_F|a,l)P(a|D,l), P_{\min}(w_F|l)\right)$$

because

$$\begin{aligned}\sum_a P_{\min}(w_F|l)P(a|D,l) &= P_{\min}(w_F|l)\sum_a P(a|D,l) \\ &= P_{\min}(w_F|l)\end{aligned}$$

we set

$$P_{\min}(w_F|l) = \frac{1}{\max_a(n(a)) + V}$$

where $\max_a(n(a))$ is the highest link count observed for an article and $V$ is the vocabulary size. We use three language models. The first, $P_{link}(Q|l,D)$, computes the likelihood of the query straighforward with the components as defined. The second, $P_{link'}(Q|l,D)$, tries to account for the training basis problem by weighting all counts in the query by multiplying them with the probability that they are linked. The third, $P_{LDA}(Q|D)$, estimates the probability according to the estimates obtained by the LDA models, in the same way as done in [Wei and Croft, 2006] for monolingual retrieval. For the LDA model we combined all runs from the cut-off corpus. We get:

$$\log P_{link}(Q|l,D) = \sum_{w\in Q} n(w,Q)\log P_{link}(w|D,l)$$

$$\log P_{link'}(Q|l,D) = \sum_{w\in Q} n(w,Q)P(l|w)\log P_{link}(w|D,l)$$

$$\log P_{LDA}(Q|D) = \sum_{w\in Q} n(w,Q)\log P_{LDA}(w|D)$$

where

$$P_{LDA}(w|D) = \psi^{(w)T}\theta^{(D)}$$

is calculated as described in section 3.2.

As a simple form of model combination, we interpolated the query log probabilities of the LDA model linearly with those of each link model. So, for instance, the combination with the unweighted model becomes:

$$\log P_\alpha(Q|D) = \alpha \log P_{link}(Q|l,D) + (1-\alpha)\log P_{LDA}(Q|D)$$

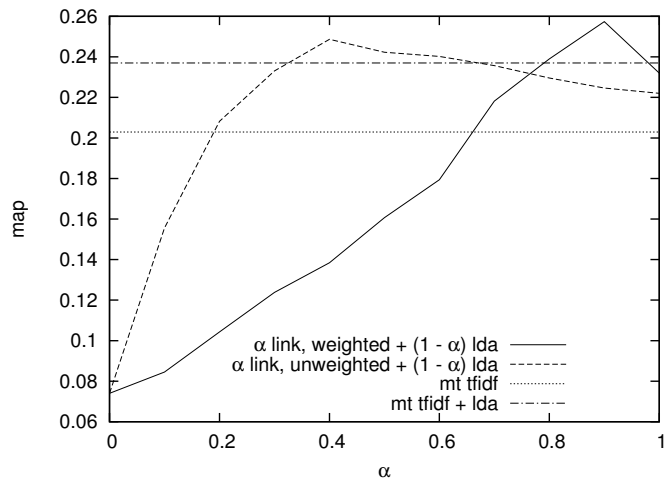

Figure 17: Combination of language models only trained on Wikipedia. Log-probabilities are interpolated on query level.

Figure 17 and Table 8 give a survey of the results. One can see that the link model alone is already better than the tf.idf baseline based on Moses and Europarl. Interestingly, the LDA model gives worse results when applied in a language modeling setup than when used based on sampling statistics on both the query and the document side, as done in chapter 4.3. This might be so because sampling takes the whole context in the query into account, which may be seen as a way of disambiguation of the query words. By the combination of the link models with LDA in the language modeling setup we achieve better *map* scores on CLEF2000 than in the best vector retrieval setup, however, the results are worse for *gmap* compared to the vector retrieval setup.

Table 8: Results for language model based retrieval (CLEF2000)

| method | combination | parameters | map | gmap |
|---|---|---|---|---|
| LDA | | cut-off, all runs | .074 | .002 |
| links, unweighted | | | .223 | .075 |
| links, weighted | | | .232 | .088 |
| LDA + links unweighted | query level | $\alpha = .4$ | .249 | .091 |
| LDA + links weighted | query level | $\alpha = .9$ | .258 | .083 |
| LDA + links | word level | $\alpha = .3, \beta = .9$ | **.303** | **.108** |

### 5.3.2 Model Combination on Word Level

The second possible form of model combination is to interpolate word distributions given a document. Here, the question arises how to weight the query words, as a weighting according to link-probability seems reasonable and turned out to be successful for the link model, but is not justified for LDA. Hence, we used two model parameters $\alpha$ and $\beta$ to interpolate distributions and weightings respectively. The query log-likelihood becomes:

$$\log P_{\alpha,\beta}(Q|D) = \sum_{w \in Q} n(w,Q) \left[\beta P(l|w) + (1-\beta)\right] \log \left[\alpha P_{link}(w|D,l) + (1-\alpha)P_{LDA}(w|D)\right]$$

The LDA-distributions are smooth by defininition, so for $0 \leq \alpha < 1$ there is no zero-probability problem. Since the smoothing method applied in the query level combination (which was necessary there) is a little ad-hoc, we decide to use no such smoothing scheme in the combination on word level. However, this keeps us from calculating values for $\alpha = 1$. We performed a grid search for the parameter values of $\alpha$ and $\beta$, using a step size of .05, and find the optimum at $\alpha = .3, \beta = .9$ with $map = .303$. This value is better than any of those reported for this track in CLEF 2000 [Braschler, 2001, Hedlund et al., 2001].

### 5.4 Discussion

We explored three ways of using translation models based on Wikipedia links and evaluated on the CLEF2000 dataset. The first is to construct a bag-of-words representation and to do retrieval with a tf.idf-scheme, the second is to compare multinomial distribution vectors with association strengths to Wikipedia articles, and the third is based on a language modeling formulation.

The pure link-based bag-of-words approach could not reach the performance of vector space retrieval done with Moses machine translation output trained on Europarl. However, in combination with LDA vectors it performed better than that and almost as well as the machine translation output in the same combination. The distribution vectors showed poor performance, however, still better than the ESA vectors on the same dataset. The language modeling approach is clearly most promising, as there is much room for the exploration of different smoothing and combination schemes. We tried two different schemes of language model interpolation with LDA, one based on combination on query level, the other based on combination on word level. While the query level approach has fewer parameters to be set, we obtained better results with the word level approach. The language modeling setting did not only outperform all our results obtained with Moses machine translation, it also produced results that compare favorably against values reported for the CLEF 2000 ad-hoc track[14].

# 6 Conclusion

In this thesis, we compare three concept models for cross-language retrieval: First, we adapt probabilistic Latent Semantic Analysis (pLSA) for multilingual documents. Experiments with different weighting schemes show that a weighting method favoring bridging documents of equal length in both language sides gives best results. Together with the finding that both monolingual and multilingual Latent Dirichlet Allocation (LDA) behave alike when applied for such documents, we use a training corpus built on Wikipedia where all documents are length-normalized. Thus, we observe a considerable improvement over previously reported scores on the Multext JOC corpus for LDA.

Another focus of our work is on model combination. For this end we include Explicit Semantic Analysis (ESA) in the experiments. We observe that apart from direct translation finding, where it had been evaluated originally, ESA is not competitive with LDA in a query based retrieval task on the CLEF 2000 data set. All concept models perform significantly worse than a model based on machine translation output. However, the combination of machine translation output with concept model information increased performance by 21.1% *map* in comparison to machine translation alone.

Machine translation relies on parallel corpora, which may not be available for many language pairs. Concept models like ESA or LDA only capture broad semantic relatedness on document level which might be to unspecific for proper names or other fine-grained distinctions. Therefore, we explore how much cross-lingual information can be carried over by a more specific information source in Wikipedia, namely linked text. We observe, that in a simple vector combination with LDA it already gives results comparable to, but not as good as the best result obtained in a combination run with machine translation.

---

[14] For licensing reasons, we are not allowed to make a direct comparison, but see [Braschler, 2001] for a survey.

The best overall results are obtained using a language modeling approach, entirely without information from parallel corpora. The language modeling approach leaves much space for further experimentation, since combination of different models is not straightforward. The need for smoothing raises interesting questions on soundness and efficiency. Link models capture only a certain kind of information and suggest weighting schemes to emphasize particular words. For a combined model, another interesting question is therefore how to integrate different weighting schemes. Using a very simple combination scheme, we obtained values that compare favorably to results reported previously on the CLEF 2000 dataset.

We have shown how probabilistic techniques that work without parallel training texts can be applied to replace other cross-language retrieval methods. In principle, our methods can be employed wherever language barriers are to be overcome on text level. After all, it would be interesting to see whether our combination techniques, successful for cross-language retrieval, can also contribute to monolingual retrieval.